\documentclass[12pt]{article}

\usepackage[DIV13]{typearea}
\usepackage{amsmath}
\usepackage{amssymb}
\usepackage{amsfonts}
\usepackage{bbold}
\usepackage{cite}
\usepackage{inputenc}
\usepackage{graphicx}
\usepackage{pstricks}
\usepackage{color}

\newcommand{\al}{\alpha}
\newcommand{\be}{\beta}

\newcommand{\om}{\omega}

\def\cL{{\cal L}}

\def\5{\overline 5}

\newcommand{\diag}{\text{diag}}
\newcommand{\hc}{\text{h.c.}}
\newcommand{\unity}{\mathbb{1}}
\newcommand{\nn}{\nonumber}
\newcommand{\mean}[1]{\langle#1\rangle}

\def\eq#1{{eq.~(\ref{#1})}}

\def\vev#1{\left\langle #1\right\rangle}

\def\hbar{\hspace{0pt}\raisebox{1pt}{$-$} \hspace{-7pt} h}

\newcommand{\beq}{\begin{equation}}
\newcommand{\eeq}{\end{equation}}
\newcommand{\bac}{\beq\begin{array}}
\newcommand{\eac}{\end{array}\eeq}
\newcommand{\ba}{\begin{array}}
\newcommand{\ea}{\end{array}}
\newcommand{\bea}{\begin{eqnarray}}
\newcommand{\eea}{\end{eqnarray}}

\begin{document}
\begin{titlepage}
\vspace*{-1cm}
\phantom{hep-ph/***}

\hfil{CFTP/09-030}
\hfil{DFPD-09/TH/14}
\hfil{IFIC/09-40}

\vskip 1.5cm

\begin{center}
{\Large\bf Tri-Bimaximal Lepton Mixing and Leptogenesis}
\end{center}

\vskip 0.5  cm

\begin{center}
{\large D. Aristizabal Sierra}~$^{a)}$\footnote{e-mail address: daristi@lnf.infn.it},
{\large F. Bazzocchi}~$^{b)}$\footnote{e-mail address: fbazzoc@few.vu.nl},
\\\vskip .2cm
{\large I. de Medeiros Varzielas}~$^{c)}$\footnote{e-mail address: ivo@cftp.ist.utl.pt},
{\large L. Merlo}~$^{d)}$\footnote{e-mail address: merlo@pd.infn.it} and {\large S. Morisi}~$^{e)}$\footnote{e-mail address: morisi@ific.uv.es}
\\
\vskip .7cm
$^{a)}$~INFN, Laboratori Nazionali di Frascati,C.P. 13, I-00044 Frascati, Italy
\\
\vskip .1cm
$^{b)}$~Department of Physics and Astronomy, Vrije Universiteit Amsterdam,\\
1081 HV Amsterdam, The Netherlands
\\
\vskip .1cm
$^{c)}$~CFTP, Departamento de F\'isica, Instituto Superior T\'ecnico,\\
Av. Rovisco Pais, 1, 1049-001 Lisboa, Portugal
\\
\vskip .1cm
$^{d)}$~Dipartimento di Fisica `G.~Galilei', Universit\`a di Padova
\\
INFN, Sezione di Padova, Via Marzolo~8, I-35131 Padua, Italy
\\
\vskip .1cm
$^{e)}$~AHEP Group, Institut de F\'isica Corpuscular --
  C.S.I.C./Universitat de Val\`encia \\
  Edificio Institutos de Paterna, Apt 22085, E--46071 Valencia, Spain
\end{center}
\vskip 0.7cm
\begin{abstract}
  In models with flavour symmetries added to the gauge group of the
  Standard Model the CP-violating asymmetry necessary for leptogenesis
  may be related with low-energy parameters. A particular case of
  interest is when the flavour symmetry produces exact Tri-Bimaximal
  lepton mixing leading to a vanishing CP-violating
  asymmetry.  In this paper we present a model-independent discussion
  that confirms this always occurs for unflavoured leptogenesis in type I
  see-saw scenarios, noting however that Tri-Bimaximal mixing does not imply a vanishing asymmetry in general
  scenarios where there is interplay between type I and other
  see-saws.  We also consider a specific model where the exact
  Tri-Bimaximal mixing is lifted by corrections that can be
  parametrised by a small number of degrees of freedom and analyse in
  detail the existing link between low and high-energy parameters -
  focusing on how the deviations from Tri-Bimaximal are connected to
  the parameters governing leptogenesis.
\end{abstract}
\end{titlepage}
\setcounter{footnote}{0}
\vskip2truecm

%
%

\section{Introduction}

Results from neutrino oscillation experiments \cite{Fukuda:1998mi}
have firmly established that neutrinos have tiny but non-zero masses.
From a theoretical perspective the smallness of neutrino masses can be
well understood within the see-saw mechanism \cite{seesaw}, in which
the Standard Model (SM) is extended by adding new heavy states. Light
neutrino masses are generated through effective operators which are
typically suppressed by the masses of the states giving rise to the
see-saw.  In type I see-saw the extra states are right-handed (RH)
neutrinos with large Majorana masses. Apart from providing an
explanation for the origin of neutrino masses, the mechanism
contains all the necessary ingredients for a dynamical generation of a
cosmic lepton asymmetry through the decays of the heavy singlet
neutrinos (leptogenesis): ($a$) Lepton number violation arising from
the Majorana mass terms of the new fermionic states; ($b$)
CP-violating sources from complex Yukawa couplings; ($c$) departure
from thermal equilibrium in the hot primeval plasma at the time the
singlet neutrinos start decaying. This lepton asymmetry is then
reprocessed into a baryon asymmetry through $B+L$ violating anomalous
electroweak processes \cite{Kuzmin:1985mm} thus yielding an
explanation to the origin of the baryon asymmetry of the Universe
\cite{Hinshaw:2008kr} i.e.  baryogenesis through leptogenesis (for a
recent review see \cite{Davidson:2008bu}).

The structure of mixing in the leptonic sector suggested by
experimental data is in sharp contrast with the small mixing that
characterises the quark sector. Observations indicate that solar
neutrino oscillation is described by a large but non-maximal mixing
angle, atmospheric neutrino oscillation is described by maximal or
nearly-maximal angle, and reactor data puts a small upper bound on the
third angle \cite{data,FogliIndication,MaltoniIndication}. This mixing
pattern is well described by the so-called Tri-Bimaximal (TB) scheme
\cite{HPS} which corresponds to a unitary matrix of the form
\begin{equation}
  U_{TB} =
  \begin{pmatrix}
    \sqrt{2/3} & 1/\sqrt3 & 0         \\
    -1/\sqrt6  & 1/\sqrt3 & -1/\sqrt2 \\
    -1/\sqrt6  & 1/\sqrt3 & +1/\sqrt2 \\
  \end{pmatrix}\,,
\end{equation}
and to the following mixing angles:
\beq
\sin^2\theta_{13}^{TB}=0\qquad\sin^2\theta_{23}^{TB}=1/2\qquad\sin^2\theta_{12}^{TB}=1/3\;.
\eeq

This particular mixing structure can be interpreted as a signal of an
underlying symmetry\footnote{for a different approach see
  \cite{AFM_BimaxS4}.} and has motivated a great deal of studies
aiming to determine the possible flavour symmetry responsible for such
a pattern. A large amount of discrete and continuous symmetries have
been considered
\cite{TBA4,continuous,others,MR_A4,BMV_A4,AF_Extra,BH_Geometric,
  Ma_TBMandSUSYwithA4,AF_Modular,HKV_A4,AFL_Orbifold,He_Renorm,
  MPT_SO10xA4,Yin_A4,Altarelli_Lectures,AFH_SU5,BMPT_SU3,AG_AF,
  HMV_Nie,LinPredictive,CDGG_Extra,JM_A4Lepto,BFM_SO10xA4,Riazuddin,
  CK_Form,Lin_Lepto,AM_Lin,BFRS_A4Lepto,HMV_ILSS,Lin_LargeReactor,FHLM_Tp,BMM_SS,BM_S4}
and among them discrete non-Abelian ones have been found to be
particularly interesting as they can more naturally lead to the required
pattern. In the realization of explicit models, a general feature is
the breaking of the flavour symmetry: this is a well known result of a
no-go theorem \cite{LV_Theorem,AF_Extra} that applies in the vast
majority of relevant cases; it could be evaded, for example
using light Higgs fields charged under the flavour symmetry, but
inconsistencies related to flavour-changing neutral current or
lepton-flavour violating processes could appear. On the contrary, allowing
only heavy Higgs fields charged under the flavour symmetry, it is
possible to avoid these dangerous effects \cite{LFV_FS}.

Global fits \cite{FogliIndication} to the data provides a subtle hint
of a deviation from the TB scheme and therefore it is desirable if the
flavour symmetry predicts TB at leading order (LO) and allows
perturbations at higher orders.
It is possible to
constrain the amount of these corrections by comparing
the TB value of the mixing angles to their
experimental measurements: the solar angle is known with the lowest
relative error and as a result it fixes the upper bound of the
deviations at about $0.05$. Avoiding any parameter tuning or
particular relations among the deviations, we expect that the other LO
mixing angles are perturbed by quantities of the same order of
magnitude: in particular the corrected $\theta_{13}$ is expected to be
non-vanishing, but very small.  \footnote{for an alternative proposal
  see \cite{Lin_LargeReactor}.}

In order to explain the baryon asymmetry of the Universe by
leptogenesis, CP violation in the leptonic sector is needed. In
principle it can be argued that leptogenesis is supported by any observation of CP violation in the
leptonic sector, e.g. in neutrino oscillation experiments. However, since generically the baryon asymmetry
is insensitive to the low energy CP-violating phases
\cite{Branco:2001pq,Davidson:2007va} a definitive conclusion can not
be established from such an observation. In contrast, in models based
on flavour symmetries that predict the TB mixing pattern, the
parameter space is further constrained and as a result one could
expect, quite generically, some link between low-energy observables
and leptogenesis.  As pointed out in \cite{JM_A4Lepto}, in the context
of an $A_4$ flavour symmetry model with type I see-saw the
CP-violating asymmetry ($\epsilon_{N_\al}$) vanishes in the limit of
exact TB mixing, with leptogenesis becoming viable only when deviations
from this pattern are taken into account.  The explicit structure of
the corrections responsible for these deviations are model-dependent
and therefore whether a connection between $\epsilon_{N_\al}$ and
low-energy parameters can be established will depend on the particular
realization.

In this paper we extend upon the work in \cite{JM_A4Lepto}. In
particular, we study the viability of leptogenesis in the context of
models based on an arbitrary flavour symmetry leading to the TB lepton
mixing pattern through the see-saw mechanism. When there is only type I see-saw and independently of the
nature of the underlying symmetry, we conclude that
$\epsilon_{N_\al}=0$ in the limit of exact TB mixing or any other exact mixing schemes where the mixing matrix consists purely of numbers - such as Bi-maximal mixing \cite{Barger:1998ta}, golden-ratio mixing \cite{Kajiyama:2007gx} and some (but not all) cases of Tri-maximal mixing \cite{Harrison:1999cf, Grimus:2008tt}. Under these conditions, only deviations from the flavour symmetry
imposed pattern yield $\epsilon_{N_\al}\neq 0$.
It is important to note that this result is not in general valid in the presence of other
types of see-saw (e.g. with the interplay of type I and type II).

Following from the model-independent proof we consider particular cases. We check our result by considering several models
discussed in the literature. Finally, we also take a specific simple $A_4$
flavour model \cite{Lin_Lepto}, where low-energy observables arising
from TB deviations can be linked to the CP-violating asymmetry in a
straightforward manner and analyse it in more detail.

Our discussion will be entirely devoted to ``unflavoured'' leptogenesis scenarios: in
the framework of flavour symmetry models predicting TB mixing the
heavy singlet neutrinos typically have masses above $10^{13}$~GeV and
for $T\gtrsim 10^{12}$ GeV lepton flavours are indistinguishable
\cite{Nardi:2006fx,Abada:2006fw}.

This paper is organised as follows: in section \ref{sec:basis} we fix
our notation and briefly comment upon some generic aspects of
leptogenesis.
For completeness of our results, in section \ref{sec:unfamiliar} we
present a brief analysis of randomly generated TB mixing and its
implications for the CP-violating asymmetry. %
We turn to the main subject of this paper in section \ref{exTBFS},
showing that an exact mixing scheme enforced by a flavour symmetry in
scenarios with just type I see-saw leads to a vanishing CP-violating
asymmetry. Leptogenesis becomes potentially viable only when
higher-order flavour symmetry corrections lift the pattern - or
otherwise if other types of see-saw (e.g. type II) are also present. %
In section \ref{sec:MBtutu} we confirm our model-independent results
in particular realizations, and in section \ref{sec:A4_2} we analyse
in detail a specific model in which low-energy parameters and the
CP-violating asymmetry are directly related in a simple way. %
Finally in section \ref{sec:end} we conclude by summarizing our
results.

%
%

\section{The basic framework}
\label{sec:basis}

In this section we will establish both the notation and a choice of a
convenient basis. Let us consider the leptonic part of the SM
Lagrangian extended with three fermionic heavy singlets $N_\alpha$
\footnote{The subsequent analysis is done for three RH neutrinos, but it can be generalised to an arbitrary number with the conclusions being independent of it.}
\beq
\label{eq:Lagrangian}
-\cL= (Y)_{ij}\overline{L}_i H\ell^c_j +
(\lambda)_{i \alpha}\overline{L}_i \widetilde{H} N_\alpha  +
\dfrac{1}{2}(M_R)_{\al \be}(N^c_\al)^T N_\be+\hc\,.
\eeq
Here $L_i$ are the lepton $SU(2)$ doublets, $\ell^c_i$ are the complex conjugate charged
lepton $SU(2)$ singlets and $H$ ($\widetilde H=i\sigma_2 H^*$) is the
Higgs $SU(2)$ doublet. Latin indices $i,j\dots$ label lepton flavour,
whereas Greek indices $\al, \be\dots$ denote RH species.
$Y$, $\lambda$ and $M_R$ are $3\times3$ matrices in flavour space.

At energy scales well below the RH neutrino masses, light neutrino masses are
generated via effective operators. The effective Majorana neutrino mass matrix is
\beq
\label{eq:lightNeu-MM}
m_\nu=-m_D \, M_R^{-1}\, m_D^T\,,
\eeq
where $m_D=\lambda\,v/\sqrt2$ ($v\simeq 246$ GeV). We then consider the unitary matrices $U_\ell$, $U_{\ell^c}$ and $U_\nu$, which diagonalise the charged lepton and neutrino mass matrices:
\beq
\hat{m}_{\ell}=U_{\ell}^\dag Y U_{\ell^c}\;\dfrac{v}{\sqrt2}\qquad\qquad \hat{m}_\nu=U_\nu^Tm_\nu U_\nu\;,
\eeq
where the ``$\hat{\phantom{b}}$'' refers to a diagonal matrix. The lepton mixing matrix is defined by
$U_\ell$ and $U_\nu$:
\beq
U=(U_\ell)^{\dagger} U_\nu\;.
\label{leptonmixing}
\eeq
From now on we will assume that in the basis in which the
charged lepton mass matrix is diagonal, $m_\nu$ is
exactly diagonalised by the TB mixing matrix $U_{TB}$ and therefore
\beq
\label{eq:diagonalization}
\hat{m}_\nu=D\,U_{TB}^T\, m_\nu\,U_{TB}\,D\,,
\eeq
where $D$ accounts for the low-energy Majorana phases
\beq
\label{eq:majoranaPhases}
D = diag (e^{i \varphi_1}, e^{i\varphi_2}, 1)\,.
\eeq
In general $m_D$ as well as $M_R$ ($M_R=M_R^T$) are complex matrices which can
be diagonalised as follows
\beq
\label{def}
\ba{rcl}
\hat{m}_D&=&U_L^\dag  \, m_D\, U_R \,,\\
\hat{M}_R&=&V_R^T \, M_R\, V_R \,,
\ea
\eeq
with $U_L,U_R,V_R$ $3\times 3$ unitary matrices,
characterised in general by 3 rotation angles and 6 phases.

According to eq. (\ref{def}) the effective neutrino mass matrix in
(\ref{eq:lightNeu-MM}) can be written as
\beq
\label{ss}
m_\nu= - U_L\,\hat{m}_D \, (U_R^\dag\,V_R)\, \hat{M}_R^{-1}\, (V_R^T U_R^*)\,\hat{m}_D\,U_L^T\,.
\eeq
The requirement of having exact TB diagonalisation can be written
either in terms of constraints over the light neutrino mass matrix
entries, namely
\beq
\label{cond1}
\ba{rcl}
m_{\nu_{12}}&=&m_{\nu_{13}}\,,\\
m_{\nu_{22}}&=&m_{\nu_{33}}\,,\\
m_{\nu_{11}}&=&m_{\nu_{22}}+m_{\nu_{23}}-m_{\nu_{12}}\,,
\ea
\eeq
or, according to eqs. (\ref{eq:diagonalization}) and (\ref{ss}), requiring that
\beq
\label{ssdiag}
\hat{m}_\nu= -  D\,(U_{TB}^T U_L)\,\hat{m}_D \, (U_R^\dag\,V_R)\,
\hat{M}_R^{-1}\, (V_R^T U_R^*)\,\hat{m}_D\,(U_L^T U_{TB})\,D
\eeq
is diagonal and real. It is useful to introduce the notation of the Dirac neutrino mass matrix in the basis in which the RH
neutrino mass matrix $\hat{M}_R$ is real and diagonal:
\beq
\label{eq:mDR_def}
m_D^R\equiv m_DV_R\,.
\eeq


\subsection{General remarks on leptogenesis}
\label{sec:gen-rem-lepto}
As mentioned in the introduction, singlet neutrinos in flavour
symmetry models typically have masses above $10^{13}$ GeV. Thus,
within these frameworks leptogenesis proceeds at temperatures at which lepton
flavour effects can be completely neglected. In the standard thermal
leptogenesis scenario singlet neutrinos $N_{\al}$ are produced by
scattering processes after inflation. Subsequent out-of-equilibrium
decays of these heavy states generate a CP-violating asymmetry given
by \cite{Davidson:2008bu,Covi:1996wh}
\begin{equation}
  \label{eq:cp-asymm}
  \epsilon_{N_\al} = \frac{1}{4v^2 \pi (m_D^{R\,\dagger} \;m_D^R)_{\al\al}}
  \sum_{\be\neq \al}{\mathbb I}\mbox{m}\left[\left((m_D^{R\,\dagger} \;m_D^R)_{\be\al}\right)^2\right]
  f(z_\be)\,,
\end{equation}
where $z_\be=M_\be^2/M_\al^2$ and the loop function can be expressed as
\begin{equation}
  \label{eq:loop-func}
  f(z_\be) =
  \sqrt{z_\be}
  \left[
    \frac{2 - z_\be}{1 - z_\be} - (1 + z_\be)\;
    \log\left(\frac{1 + z_\be}{z_\be}\right)
  \right]\,.
\end{equation}
Depending on the singlet neutrino mass spectrum the loop function
can be further simplified. In the hierarchical limit ($M_\al\ll M_\be$)
this function becomes
\begin{equation}
  \label{eq:loop-function-HS}
  f(z_\be) \to -\frac{3}{2\sqrt{z_\be}}\,,
\end{equation}
whereas in the case of an almost degenerate heavy neutrino spectrum
($z_\be=1+\delta_\be$, $\delta_\be\ll 1$) it can be rewritten as
\begin{equation}
  \label{eq:loop-function-DS}
  f(1+\delta_\be)\simeq -\frac{1}{\delta_\be}\,.
\end{equation}
In any case, as can be seen from eq. (\ref{eq:cp-asymm}), whether the
CP-violating asymmetry vanishes will be determined by the Yukawa
coupling combination $m_D^{R\,\dagger} m_D^R$.

%
%

\section{CP asymmetry  and exact TB mixing without any underlying  flavour symmetry}
\label{sec:unfamiliar}

While the TB mixing pattern can be well understood as a consequence of
an underlying flavour symmetry, in principle it might be that it arises from a random set of parameters (though quite unlikely).  For completeness, in this section we consider this
possibility and study the consequences on the
CP-violating asymmetry.  Neutrino mixing angles are fixed to
satisfy the TB mixing pattern and in addition to the measured mass
squared differences we have a set of eight constraints on the parameter space: the TB mixing condition enforces the
relations in eq. (\ref{cond1}), yielding six constraints (from
the real and imaginary parts of the mass matrix entries); the atmospheric
and solar mass scales provide the remaining two.

To determine the effect of such constraints on
$\epsilon_{N_\al}$ it is practical to use a parametrisation of
$m_D$ that ensures that the TB mixing and the correct neutrino
masses are obtained. In the basis in which the RH neutrino mass matrix
is diagonal and real it is convenient to introduce the orthogonal
complex matrix $R$ defined by the so-called Casas-Ibarra
parametrisation \cite{Casas:2001sr}, namely
\beq
  \label{eq:casas-ibarra}
R^*= (\hat{m}_\nu)^{-1/2} \,U^T\,m_D^R\, (\hat{M}_R)^{-1/2}\,.
\eeq
All low-energy observables are contained in the leptonic mixing matrix
$U$ and in the diagonal and real light neutrino mass matrix $\hat{m}_\nu$. The matrix $R$
turns out to be very useful in expressing the CP-violating asymmetry
parameter. Considering for simplicity the case of
hierarchical RH neutrinos ($M_1 \ll M_2 \ll M_3$ - thus validating the
approximation in \eq{eq:loop-function-HS}), \eq{eq:cp-asymm} can
be rewritten as
\begin{equation}
  \label{eq:cp-asymm-CI}
  \epsilon_{N_\al} = -\frac{3 M_\al}{8 \pi v^2}
  \frac{{\mathbb I}\mbox{m}
    \left[\sum_j m_j^2 R_{j\al}^2\right]}
  {\sum_j m_j |R_{j\al}|^2}\,,
\end{equation}
where $m_j\equiv(\hat m_\nu)_{jj}$.
Once the RH neutrino mass spectrum and low-energy observables are
fixed, random values of $m^R_D$ correspond to random values of $R$.
It is shown by eq. (\ref{eq:cp-asymm-CI}) that leptogenesis is completely insensitive to low-energy lepton mixing and CP-violating phases
\cite{Branco:2001pq} \footnote{This statement is in general also true
  in flavoured leptogenesis \cite{Davidson:2007va}.} and therefore
the viability of leptogenesis is not at all related with any accidental mixing pattern considered. The
CP-violating asymmetry is determined by the values of the entries of
$R$ which are arbitrary in the absence of any flavour
symmetry, and consequently $\epsilon_{N_a}\neq 0$ in general and its absolute
value depends upon the heavy fermionic singlet masses, the light
neutrino masses and $R$.

\begin{figure}[t]
  \centering
  \includegraphics[width=10cm,height=7cm]{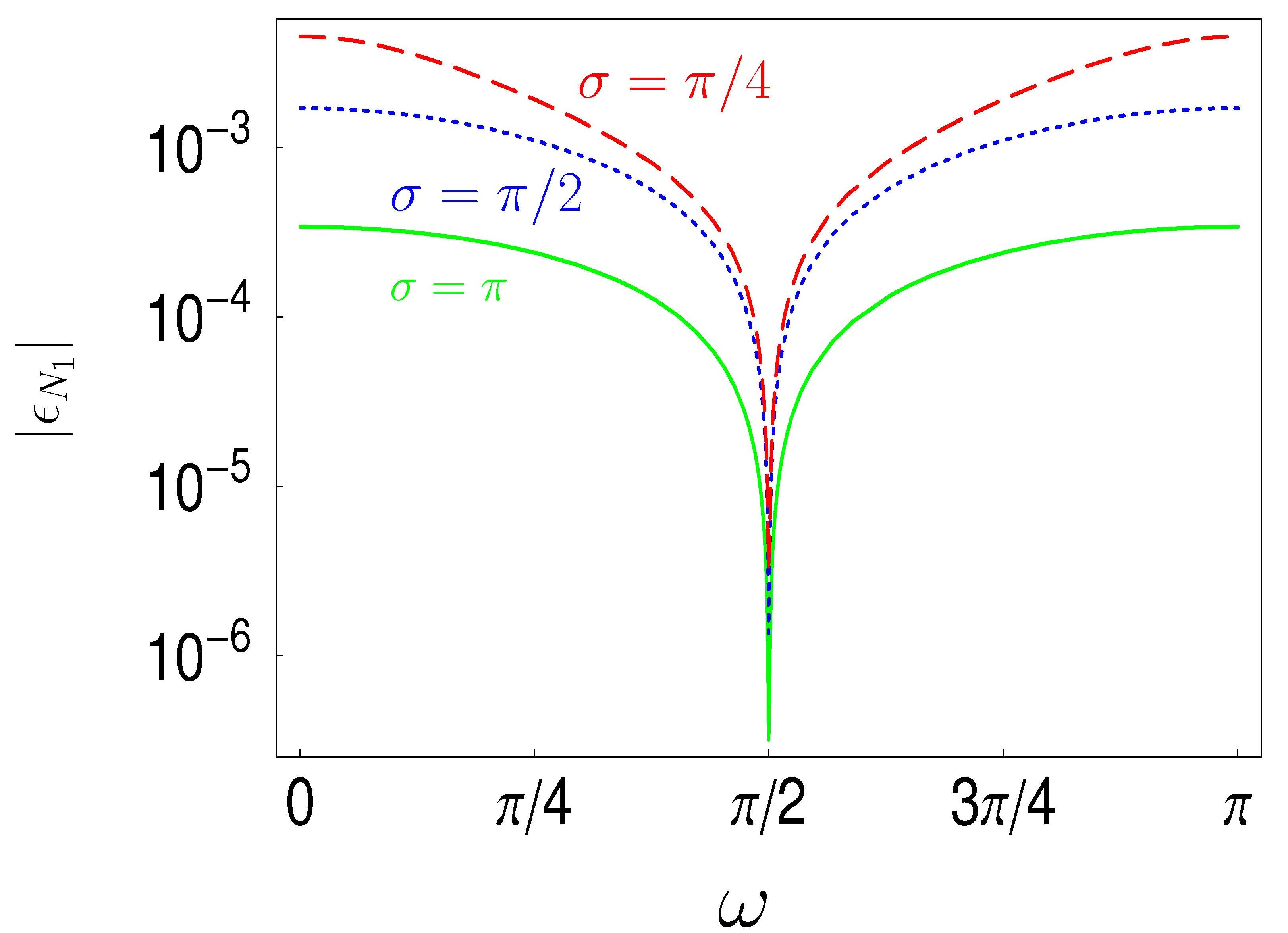}
  \caption{CP-violating asymmetry as a function of the angle $\omega$
    for different values of $\sigma$. $M_1$ is fixed to
    $10^{13}$ GeV and $\Delta m^2_{\text{atm}}$ to $2.39\times10^{-3}\;\mathrm{eV}^2$
    \cite{MaltoniIndication} (see the text for further details).}
  \label{fig:cp-asymm}
\end{figure}

To illustrate this point we consider the
case in which only $N_1$ decays are relevant for the generation of a
lepton asymmetry. We assume normal hierarchy for the light
neutrino spectrum and a simple $R=R_{13}(\rho_{13})$ with $\rho_{13}=\omega + i \sigma$ (i.e. $R$ is a $\rho_{13}$
rotation matrix). Under these assumptions
the CP-violating asymmetry in \eq{eq:cp-asymm-CI} becomes
\begin{equation}
  \label{eq:cp-asymm-simpl}
  \epsilon_{N_1} = -\frac{3 M_1 \sqrt{\Delta m^2_{\text{atm}}}}{2 \pi v^2}
    \frac{\cos \omega\; \sinh \sigma}{\sqrt{\cosh 2\sigma - \cos 2\omega}}\,.
\end{equation}
From figure \ref{fig:cp-asymm} it can be seen that barring the cases
$\omega=\pi/2$ and/or $\sigma=0$ the CP-violating asymmetry does not
vanish and its values are well within the range required for successful
leptogenesis, regardless of the mixing pattern.

%
%

\section{Implications of flavour symmetries on the CP asymmetry}
\label{exTBFS}

We consider now the case in which an underlying flavour symmetry
enforces an exact mixing pattern. It will be evident throughout the proof that it holds for any mixing pattern where the mixing matrix consists purely of numbers, but we will assume TB mixing for definiteness.

Within the case considered the transformation
properties of $L_i$ and $N_\alpha$ under the flavour symmetry group
($G_f$) determine the structure of $m_D$ and $M_R$ (which are no longer arbitrary).
Indeed, these matrices can be regarded as form-diagonalisable
matrices \cite{LV_Theorem}, i.e. the
parameters which determine their eigenvalues are completely
independent from the parameters that define their diagonalising
matrices. Accordingly, vanishing off-diagonal elements of $\hat{m}_\nu$ in
eq. (\ref{ssdiag}) can arise only if
\begin{equation}
\label{eq:rot-mat-relations}
U_{TB}^T U_L= P_L \, O_{D_i} \quad \mbox{and}\quad
U_R^\dag\,V_R=O^\dag_{D_i}\,P_R\,O_{R_{m}}\,,
\end{equation}
where $P_{L,R}=\text{diag}(e^{i\alpha^{R,L}_1},e^{i\alpha^{R,L}_2},
e^{i \alpha^{R,L}_3})$ whereas $O_{D_i}$ and $O_{R_m}$ are respectively unitary
and orthogonal matrices that arbitrarily rotate the
$i$ and $m$ degenerate eigenvalues of $m_D$ and $M_R$ such that if
$m_D$ ($M_R$) has no degenerate eigenvalues $O_{D_i}=\unity$
($O_{R_m}=\unity$). Note that the requirement of having canonical kinetic terms
in addition to preserving the $m$-fold degeneracy of the RH neutrino mass matrix
enforce $O_{R_m}$ to be real. Although $O_{D_i}$ and $O_{R_m}$ do not have any
effect in eq. (\ref{ssdiag}) they do affect the structure of
$U_{L,R}$ and $V_R$ and correspondingly of $m_D$ (see eq.
\eqref{def}). $V_R$ can be defined in such a way that $\hat{M}_R$ is real, and the phases contained in $\hat{m}_D$ are now
denoted by $\gamma_i$ and must obey: $\varphi_i + \alpha^{R}_i + \alpha^{L}_i + \gamma_i= 2 k \pi$
and $\alpha^{R}_3 + \alpha^{L}_3 + \gamma_3 = 2 n \pi$. It is easy to understand the conditions given in \eq{eq:rot-mat-relations} by the use of a \emph{reductio ad absurdum}. Let us consider for simplicity the case without any degeneracy in the
eigenvalues of $\hat{m}_D$ and $\hat{M}_R$: $O_{D_i}=\unity$ and
$O_{R_m}=\unity$. If the products $U_{TB}^TU_L$ and $U_R^\dag V_R$ are not
diagonal, but simply unitary matrices with non-vanishing off-diagonal entries,
then the right-hand side of eq. (12) is in general a matrix whose entries are
linear combinations of the mass eigenvalues of $\hat{m}_D$ and of $\hat{M}_R$.
In order to have $\hat{m}_\nu$ diagonal, the off-diagonal entries must vanish
and this is possible only if the respective linear combinations cancel out.
However, there are no apriori reasons to have such cancellations, since it
corresponds to have well-defined relationships between the eigenvalues of
$\hat{m}_D$ and of $\hat{M}_R$, which is, in other words, a fine-tuning.
Avoiding this possibility, the only solution is to consider \eq{eq:rot-mat-relations}.

It is useful to classify the number of degenerate eigenvalues of $m_D$
and $M_R$. There are nine cases in total: 3 for $ m_D$ ($i$=1, 2
or 3-fold degeneracy) and 3 for ${M}_R$ ($m$= 1, 2 or 3-fold
degeneracy).  In the following we will identify each case by $(i,m)$.
The cases $(3,3)$, $(2,3)$ and $(3,2)$ are not consistent with
experimental data on neutrino mass splittings, so we are left with six
viable cases:

\begin{itemize}
\item[a)]~~$(1,1)$: ${m}_D$ and ${M}_R$ have no
  degenerate eigenvalues;
\item[b)]~~$(2,1)$: ${m}_D$ with 2 degenerate eigenvalues;
\item[c)]~~$(1,2)$: ${M}_R$ with 2 degenerate eigenvalues;
\item[d)]~~$(2,2)$: ${m}_D$ and ${M}_R$ with 2 degenerate
  eigenvalues;
\item[e)]~~$(3,1)$: ${m}_{D}$ with 3 degenerate eigenvalues;
\item[f)]~~$(1,3)$: ${M}_{R}$ with 3 degenerate eigenvalues.
\end{itemize}
We proceed to show that all the viable cases obey a common
expression. In the basis in which the RH neutrinos are diagonal we
use $m_D^R$ (see \eq{eq:mDR_def}) and write
$\hat{m}_D=\delta_i\,diag(v_1,v_2,v_3)$, where we have schematically
indicated with $\delta_i$ the fact that $i$ values of
$diag(v_1,v_2,v_3)$ are equal. In other words for $\delta_i=\delta_3$
we have $diag(v_1,v_1,v_1)$ and for $\delta_i=\delta_2$ we have
$diag(v_1,v_2,v_1)$ or one of its possible permutations.
We thus have
\beq
m_D^R= U_{TB}\,P_L\, O_{D_i}\, \delta_i\,diag
(v_1,v_2,v_3)\, O_{D_i}^\dag\,P_R\, O_{R_m}\,.
\eeq
It is clear that in the subspace of the $i$ degenerate eigenvalues the
rotation $O_{D_i}$ acts as $O_{D_i}\, \delta_i\,diag (v_1,v_2,v_3)\,
O_{D_i}^\dag \to \delta_i\,diag (v_1,v_2,v_3)\,.$ Therefore we simplify the expression of $m_D^R$:
\beq
\label{vCI}
m_D^R= U_{TB}\,P_L\, \delta_i\,diag (v_1,v_2,v_3)\,\,P_R\, O_{R_m}\,.
\eeq
The next step consists in the redefinition of the $v_i$ by
absorbing $P_L,P_R$. In this way the degeneracy of the $i$ eigenvalues is broken and we finally get
\beq
\label{gen}
\ba{rcl}
m^R_D&=& U_{TB}\, diag (v_1,v_2,v_3)\,O_{R_m}\\
&=&\left( \begin{array}{ccc} \sqrt{\frac{2}{3}}v_1&\frac{v_2}{\sqrt{3}}&0\\
 -\frac{v_1}{ \sqrt{6}} &\frac{v_2}{\sqrt{3}}&-\frac{v_3}{\sqrt{2}}\\
  -\frac{v_1}{ \sqrt{6}} &\frac{v_2}{\sqrt{3}}&\frac{v_3}{\sqrt{2}}\\
  \end{array}\right)\, O_{R_m}\,.
\ea
\eeq
According to our formalism, the RH neutrino mass matrix
is trivially given by
\beq
\hat{M}_R=\delta_m \,diag(M_1,M_2,M_3)\,,
\eeq
where $\delta_m$ indicates that $m$ eigenvalues of $diag(M_1,M_2,M_3)$
are degenerate.

We now rewrite \eq{vCI}  according to the following parametrisation
\beq
\label{vCI1}
m_D^R= U_{TB}\,P\, \hat{v}\, O_{R_m}\, ,
\eeq
with $\hat{v}= diag(|v_1|,|v_2|,|v_3|)$ and all the phases absorbed in the diagonal  unitary matrix $P$.  In this basis and using the  parametrisation given in \eq{vCI1}  for $m_D^R$,  the type I see-saw formula of \eq{ssdiag} is written as
\bea
\label{vCI2}
\hat{m}_\nu &=&- D\, U_{TB}^T\, ( U_{TB} \,P\, \hat{v}\, O_{R_m}\,)\hat{M}_R^{-1}  (O^T_{R_m}\,\hat{v} \,P\, U_{TB}^T) U_{TB} \,D\nn\\
&=& (D\, P \, e^{i \pi/2})\, \hat{v} \hat{M}_R^{-1} \hat{v}\,(e^{i \pi/2} \,P \,D)=  (\hat{v} \hat{M}_R^{-1/2}  R^\dag)(R^*  \hat{M}_R^{-1/2} \hat{v} )\,,
\eea
where $D=P^*\, \, e^{-i \pi/2}$ is a consequence of our definition of $\hat{m}_\nu$ in  \eq{ssdiag}, and where we have introduced the arbitrary orthogonal complex matrix  $R$ in the last part of \eq{vCI2}. $O_{R_m}$ acts only in the subspace of the degenerate right handed neutrinos and in this subspace we have by definition $ O_{R_m} \,O^T_{R_m}= \unity$.
From \eq{vCI2} we have that
\beq
\hat{m}_\nu^{-1/2} \, \hat{v} \,\hat{M}_R^{-1/2}  R^\dag= \unity \,,
\eeq
and remembering that $R^\dag R^*= R^T R=\unity$ we arrive at our parametrisation for $R^*$
\beq
\label{ourR}
R^*=\hat{m}_\nu^{-1/2}\,  \hat{v}\, \hat{M}_R^{-1/2} \,.
\eeq
By comparing \eq{ourR} with the Casas-Ibarra parametrisation given in
\eq{eq:casas-ibarra} we deduce that in the case of exact TB mixing the
matrix $R$ is real and according to \eq{eq:cp-asymm-CI} the
CP-violating asymmetry vanishes.

Note that so far we did not refer to any specific model
realisation and we have assumed just exact TB diagonalisation of
$m_\nu$ within the context of type I see-saw. We not only confirm the
result in \cite{JM_A4Lepto} (in which a model with the $A_4$ flavour
symmetry has exact TB mixing leading to a vanishing CP-violating
asymmetry), but also extend it to any possible flavour symmetry
responsible for the exact TB scheme \footnote{This result is
  basis independent and thus remains true even assuming a non-diagonal
  charged lepton mass matrix.} .

It is also straightforward to check by replacing $U_{TB}$ with the appropriate mixing matrix that the matrix $R$ still turns out to be real for other exact mixing schemes as long as their mixing matrix also consists purely of numbers (e.g. the
corresponding matrix for the Bi-maximal mixing scheme).
Note also that although we have only considered three RH neutrinos our result is
absolutely generalisable to models with either two RH neutrinos or
more than three such as \cite{moreNR}.

\begin{figure}[t]
  \centering
  \includegraphics[width=3.5cm,height=3cm]{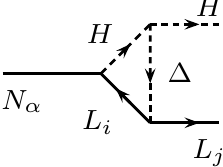}
  \caption{Vertex correction involving a triplet scalar $\Delta$.}
  \label{fig:Delta}
\end{figure}

The proof does not hold however in the presence of additional degrees of
freedom, e.g. in models involving type I and type II see-saw. Other contributions to the CP-violating asymmetry
will in general not vanish in the limit of exact TB mixing, rendering our result
invalid for situations which do not have only type I see-saw.
In scenarios with type II see-saw the details concerning
the generation of the lepton asymmetry will depend upon the
hierarchies between the triplet ($\Delta$) and the lightest RH
neutrino masses \cite{Hambye:2003ka, Antusch:2004xy}. Even in the case
$M_\Delta > M_{N_\alpha}$ ($N_\alpha$ being the lightest RH neutrino)
the CP asymmetry will receive an extra contribution from the loop
diagram shown in figure \ref{fig:Delta}. This contribution will not necessarily vanish, although it is constrained by the
TB mixing pattern \cite{future}.

An important consequence of our proof is that if the TB mixing
pattern is due to any underlying flavour symmetry in a type I see-saw
scenario, the viability of leptogenesis depends upon possible
departures from the exact pattern. In the context of models based on
discrete flavour symmetries that predict TB mixing at LO this is
achieved through next to LO (NLO) corrections. Since the size of the
deviations from TB mixing are not arbitrary, in principle one might
expect the CP-violating asymmetry to be constrained by low-energy
observables such as $\theta_{13}$ and/or the CP-violating phases.

In order to see if this is the case let us consider the most
generic situation, in which NLO corrections affect $m_\ell$, $m_D$ and
$M_R$. We can perform a linear expansion in the corrections that
appear at NLO. First, we note that $m_\ell$ is no longer diagonal
and thus we have to move to the basis in which the charged lepton
mass matrix is diagonal:
\beq
U_\ell^\dag \, m_\ell m_\ell^\dag\, U_\ell= \left(m_\ell m_\ell^\dag\right)_{diag}\;,
\eeq
where $U_\ell=\unity+U_\ell^{(1)}$, with $U_\ell^{(1)}$ the matrix of the NLO shifts.
Eq. (\ref{def}) is modified as follows
\bea
\label{def2}
\left(\unity+U_L^{(1)\dag}\right) U_L^\dag\left(m_D+m^{(1)}_D\right)U_R\left(\unity+U_R^{(1)}\right) &\simeq&
\hat{m}_D+ U_L^\dag m_D^{(1)} U_R+ U_L^{(1) \dag} \hat{m}_D+ \hat{m}_D U_R^{(1)}\nn\\
&\equiv&\hat{m}^\prime_D=\hat{m}_D +\hat{m}_D^{(1)}\,,\nn\\
\\
\left(\unity+V_R^{(1) T}\right)V_R^T \left( M_R+M^{(1)}_R \right) V_R\left(\unity+V_R^{(1)}\right)&\simeq &
\hat{M}_R+ V_R^TM^{(1)}_R V_R+V_R^{(1) T}\hat{M}_R+\hat{M}_RV_R^{(1)}\nn\\
&=& \hat{M}^\prime_R= \hat{M}_R+ \hat{M}^{(1)}_R\,.\nn
\eea
Here the unitary matrices are parametrised as the LO terms shifted by
the NLO ones. The superscript `` $^{(1)}$ '' refers to the NLO
corrections and `` $^\prime$ '' to the complete mass matrices up to
NLO. The corresponding shifts on the light neutrino masses due to the
NLO corrections can be estimated according to
\beq \mathcal{O}(\hat{m}^{\prime}_\nu-\hat{m}_\nu) \sim \mathcal{O}
(\hat{m}_D \hat{m}_{D}^{(1)}/\hat{M}_R)\sim \mathcal{O}
(\hat{m}_D^2 \hat{M}_{R}^{(1)} /\hat{M}^2_R)\;.
\eeq
Similarly, we can parametrise the shift from the exact TB pattern in the
neutrino mixing matrix:
\beq
U_\nu= U_{TB}\left(\unity+U_{TB}^{(1)}\right) D\;,
\eeq
where $U_{TB}^{(1)}$ arises by the interplay between all the
corrections. When we constrain the entries of $U_{TB}^{(1)}$ by
neutrino experimental data, we obtain constraints on $U_\ell^{(1)}$,
$U_L^{(1)}$, $U_R^{(1)}$, $V_R^{(1)}$. Experimental data on
neutrino mass splittings constrains $m_D^{(1)}$ and $M_R^{(1)}$.

We write now \eq{gen} in the new basis in which the RH neutrinos and the charged leptons are diagonal:
\beq
\label{genper}
\ba{rcl}
m_D^{R\prime}&=&\left(\unity+U_\ell^{(1)\dag}\right)U_L\left(\unity+U_L^{(1)}\right)\left(\hat{m}_D+\hat m^{(1)}_D\right) \left(\unity+U_R^{(1)\dag}\right)U_R^\dag V_R\left(\unity+V_R^{(1)}\right)\\
\\
&=&m_D^R+U_\ell^{(1)\dag}m_D^R+U_LU_L^{(1)}\hat{m}_DU_R^{\dag}V_R+U_L\hat{m}_D^\prime
U_R^\dag V_R+U_L\hat{m}_DU_R^{(1)\dag}U_R^{\dag}V_R+m_D^RV_R^{(1)}\,.
\ea
\eeq
Thus after including NLO corrections the quantity relevant for leptogenesis becomes
\beq
\ba{rcl}
m_D^{R\prime\dag} m_D^{R\prime}&=&m_D^{R\dag} m_D^R+\Bigg[m_D^{R\dag}\Bigg(U_\ell^{(1)\dag}m_D^R+U_LU_L^{(1)}\hat{m}_DU_R^{\dag}V_R+U_L\hat{m}_D^\prime
U_R^\dag V_R+\\
&&\hspace{5cm}+U_L\hat{m}_DU_R^{(1)\dag}U_R^{\dag}V_R+m_D^RV_R^{(1)}\Bigg)+\mathrm{h.c.}\Bigg]\,.
\ea
\eeq
Some comments are in order concerning this expression. The combination
$m_D^{R\dagger} m_D^R$ is shifted by NLO corrections, and in general
it is no longer real - leading to $\epsilon_{N_\alpha}\neq 0$ and
enabling viable leptogenesis. The combination of NLO corrections that
defines the shift is not directly related with any low-energy
observable. Consequently, while we conclude that general model-independent NLO corrections guarantee a
non-vanishing CP-violating asymmetry, correlations among
low-energy observables in the leptonic sector and
$\epsilon_{N_\alpha}$ can not be established unless the nature of the
corrections is well known i.e.  once the flavour model realisation has
been specified.


\section{Model building realisations of the different possibilities}
\label{sec:MBtutu}

In the previous section we have presented a model-independent proof: exact TB mixing produced by any flavour symmetry in a type I see-saw scenario corresponds to vanishing CP-asymmetry. In this section we gather the different models studied in literature which fall under the validity of the proof, and verify that they correspond to one of the six viable cases of section \ref{exTBFS}. We have also present a toy model exemplifying the $(2,2)$ case (i.e. both matrices have two degenerate eigenvalues) which has not been studied yet. We show that all models lead to a vanishing CP-asymmetry and thus this analysis serves as an ample set of examples of the validity our model-independent proof.

Before describing the flavour models proposed in the
literature, it is useful to explain the generic approach considered in flavour symmetry model building.
The main goal of these models is to explain the fermion mass hierarchies and mixing angles. To do so, an
horizontal flavour group $G_f$ is added to the gauge group of the SM
and the SM fields transform in a non-trivial way under $G_f$. Extra fields (flavons) are added to the
particle spectrum: the flavons are invariant under $SU(3)\times
SU(2)\times U(1)$, but not under $G_f$; they can acquire a non-vanishing
vacuum expectation value (VEV) which spontaneously breaks the flavour
symmetry in a well determined breaking chain. It is through the specific realisation of the breaking chain that one can achieve the goal of explaining fermion data: for
example, the lepton mixing matrix becomes the TB structure when $G_f$ is broken down to two
distinct and specific subgroups, $G_\ell$ in the charged lepton sector
and $G_\nu$ in the neutrino one, with the type of these subgroups defining the flavour structure of the mass matrices for the
leptons (which is model-dependent).

In the following analysis we specify only which $G_f$ was used, and the resulting neutrino mass matrices. We leave all other details to the original papers.


\subsubsection*{a)~~$\mathbf{(i,m)=(1,1)}$}

There are only a few examples of this case in literature. This case is particularly attractive within the context of a Grand Unified Theory (GUT). In some cases the models do not have exact TB only because they account simultaneously for the quark sector \cite{continuous}, with the Cabibbo angle generating LO deviations from exact leptonic TB - therefore they are not as interesting for our current purpose, and in \cite{Antusch:2006cw} leptogenesis within the sequential dominance framework was considered in detail (note that there is no inconsistency with our model-independent proof). Here we consider instead two other cases explicitly.
\begin{enumerate}
\item In \cite{BMPT_SU3} the authors present a model in the context of the $SO(10)$ GUT with the addition of the flavour group $G_f=SU(3)\times U(1)$. The breaking of $G_f$ down to the discrete non-Abelian group $A_4$ provides the TB pattern for the lepton mixing matrix. The neutrino mass matrices have the flavour structure:
    \beq
    m_D\propto\left(
          \begin{array}{ccc}
            A & B & 0 \\
            B & \om A & 0 \\
            0 & 0 & \om^2 A \\
          \end{array}
        \right)\quad\quad\text{and}\quad\quad
    M_R\propto\left(
      \begin{array}{ccc}
        A' & B' & 0 \\
        B' & \om A' & 0 \\
        0 & 0 & \om^2 A' \\
      \end{array}
    \right)
    \eeq
    where $\om=e^{\frac{2i\pi}{3}}$. It is straightforward to show how the correct mixing pattern is recovered by the diagonalisation of the charged lepton mass matrix and we refer to the original paper for the details. For leptogenesis what is relevant are the imaginary parts of the off-diagonal entries of the product $m_D^{R\,\dag} m_D^R$, and in this case it is a diagonal matrix.
\item Another pattern has been presented in \cite{BFM_SO10xA4} in the context of an $SO(10)$ GUT model with $A_4$ as the additional flavour group. The mass matrices have the following structure
    \beq
    m_D\propto\left(
      \begin{array}{ccc}
        A & 0 & B \\
        0 & C & 0 \\
        B & 0 & A \\
      \end{array}
    \right)\qquad\qquad\text{and}\qquad\qquad
    M_R\propto\left(
      \begin{array}{ccc}
        A' & 0 & B' \\
        0 & C' & 0 \\
        B' & 0 & A' \\
      \end{array}
    \right)\;.
    \eeq
    After considering the charged leptons the TB mixing scheme is obtained. Computing $m_D^{R\,\dag} m_D^R$, we find that the off-diagonal entries are real.
\end{enumerate}


\subsubsection*{b)~~$\mathbf{(i,m)=(2,1)}$}

There are several papers in which the Dirac neutrino mass matrix has only two independent mass eigenvalues: we can divide the discussion in terms of the flavour patterns used for the mass matrices.
\begin{enumerate}
  \item
    The first pattern is present in \cite{AF_Modular,AG_AF,JM_A4Lepto,Lin_Lepto,AM_Lin,FHLM_Tp,Lin_LargeReactor}. In the basis of diagonal charged leptons, the neutrino mass matrices have the structure:
    \beq
    m_D\propto\left(
          \begin{array}{ccc}
            1 & 0 & 0 \\
            0 & 0 & 1 \\
            0 & 1 & 0 \\
          \end{array}
        \right)\quad\quad\text{and}\quad\quad
    M_R\propto\left(
          \begin{array}{ccc}
            A'+2B' & -B' & -B' \\
            -B' & 2B' & A'-B' \\
            -B' & A'-B' & 2B' \\
           \end{array}
         \right)\;,
    \eeq
    where $M_R$ is exactly diagonalisable by the TB mixing. The product $m_D^{R\,\dag} m_D^R$ is proportional to the identity.\\
    Two different discrete groups have been used: $A_4$ in \cite{AF_Modular,AG_AF,JM_A4Lepto,Lin_Lepto,AM_Lin,Lin_LargeReactor} and $T'$ in \cite{FHLM_Tp}.

  \item
    The other pattern has been presented in \cite{BMM_SS} where the authors have used the $S_4$ discrete symmetry and it differs from the previous one in the explicit form of the Majorana mass matrix:
    \beq
    m_D\propto\left(
          \begin{array}{ccc}
            1 & 0 & 0 \\
            0 & 0 & 1 \\
            0 & 1 & 0 \\
          \end{array}
        \right)\quad\quad\text{and}\quad\quad
    M_R\propto\left(
          \begin{array}{ccc}
            2A' & B'-A' & B'-A' \\
            B'-A' & 2A'+B' & -A' \\
            B'-A' & -A' & 2A'+B' \\
           \end{array}
         \right)\;.
    \eeq
    This pattern corresponds to a completely different neutrino oscillation phenomenology, but the contribution to leptogenesis is still vanishing in the limit of exact TB mixing.
\end{enumerate}


\subsubsection*{c)~~$\mathbf{(i,m)=(1,2)}$}

There is only one pattern within this case \cite{CK_Form}. The discrete group $A_4$ is used to construct a Majorana mass matrix with two degenerate eigenvalues and a Dirac mass matrix of the TB-type.\footnote{We underline the absence of a relevant contribution to the Dirac mass matrix, the antisymmetric contraction of the two triplets in a singlet \cite{HMV_Nie}. In order to recover the TB pattern it is possible to either assume a fine-tuning on the parameters or alternatively to adapt the model to use another discrete group such as $S_4$, in which case this problem is naturally solved by its properties.}
The mass matrices are given by:
\beq
m_D\propto\left(
          \begin{array}{ccc}
            2A+B & -A & -A \\
            -A & 2A & B-A \\
            -A & B-A & 2A \\
           \end{array}
         \right)\qquad\qquad\text{and}\qquad\qquad
M_R\propto\left(
  \begin{array}{ccc}
    1 & 0 & 0 \\
    0 & 0 & 1 \\
    0 & 1 & 0 \\
  \end{array}
\right)\;.
\eeq
The product $m_D^\dag m_D$ is diagonalised by the TB mixing matrix and it is easy to verify that also the light neutrino mass matrix has this property. $m_D^{R\,\dag} m_D^R$ does not present any imaginary off-diagonal factor.


\subsubsection*{d)~~$\mathbf{(i,m)=(2,2)}$}

There are no models of this kind in the literature. The difficulty consists in the possibility that the degenerate eigenvalues of the Dirac and Majorana matrices conspire to give a degenerate light neutrino spectrum. A fully developed model is beyond the scope of this paper, but we present here an example. Although it requires some \emph{ad hoc} conditions it is sufficient to illustrate a possible setting in which both non-degenerate light neutrino spectrum and TB mixing are achieved.

The flavour group consists of $SO(3)$ (or a subgroup with an irreducible triplet representation). The additional scalar content is a set of four flavon triplets, $\phi_{123}$, $\phi_{23}$, $\phi_{2}$ and $\phi_{3}$ which get non-vanishing VEVs. At this level we fix only the VEVs of the first two flavons in such a way that $\mean{\phi_{123}}\propto (1,1,1)$ and $\mean{\phi_{23}}\propto (0,1,-1)$ (these VEVs must be orthogonal). The structure is reminiscent of the models in \cite{continuous}.

The left and RH neutrinos transform as triplets under $SO(3)$.
We assume that any additional symmetry allows the Dirac terms
\beq
(\phi_{123 i} \nu_i)(\phi_{2\alpha} N_\alpha)\;,\qquad(\phi_{23 i} \nu_i) (\phi_{3 \alpha} N_\alpha)
\eeq
and the Majorana terms
\beq
N_\alpha N_\alpha\;,\qquad (\phi_{3 \alpha} N_\alpha) (\phi_{3 \beta} N_\beta) \;.
\eeq
The term $N_\alpha N_\alpha$ by itself would lead to degenerate masses in the Majorana matrix. The degeneracy is lifted only for one of the states by the VEV $\langle \phi_3 \rangle \propto (0,0,1)$ (two eigenvalues remain degenerate). Thus the RH neutrino mass matrix has structure:
\beq
M_R \propto \left(
  \begin{array}{ccc}
    1 & 0 & 0 \\
    0 & 1 & 0 \\
   0 & 0 & x \\
  \end{array}
\right)\;,
\eeq
where $x$ parametrises that the entry receives contribution due to $\langle \phi_3 \rangle$.
In the Dirac sector one of the eigenvalues is zero. For a non-trivial choice of parameters we end up with exactly two non-zero degenerate eigenstates. With $\mean{\phi_2} \propto (0,1,0)$ and through the type I see-saw, the term $(\phi_{123 i} \nu_i)(\phi_{2\alpha} N_\alpha)$ will give rise to the solar eigenstate and the term $(\phi_{23 i} \nu_i) (\phi_{3 \alpha} N_\alpha)$ will give rise to the atmospheric eigenstate. In this case the Dirac mass matrix is:
\beq
m_D \propto \left(
  \begin{array}{ccc}
    0 & t & 0 \\
    0 & t & b \\
    0 & t & -b \\
  \end{array}
\right)\;,
\eeq
where $t$ and $b$ parametrise the contributions of $(\phi_{123} \nu) (\phi_{2} N)$ and $(\phi_{23} \nu) (\phi_{3} N)$ respectively. The effective neutrino mass matrix is diagonalised by TB mixing, as this model fits within the framework described in \cite{continuous}. There is sufficient freedom to fit the squared mass differences (as required by phenomenology), although only strongly hierarchical cases are possible due to the vanishing eigenvalue of $m_D$. The Dirac matrix has two degenerate masses by requiring $3 t^2 = 2 b^2$ (completely \textit{ad hoc}, as it requires the conspiracy of the VEVs of the flavons - we can express it as a very specific requirement on the magnitude of $\mean{\phi_2}$). It is straightforward to see that $m_D^{R\,\dag}m_D^R$ is a diagonal matrix, leading to vanishing leptogenesis.


\subsubsection*{e)~~$\mathbf{(i,m)=(3,1)}$}
This case is the most studied in literature and there are some interesting flavour patterns.
\begin{enumerate}
  \item
    The first pattern has been presented in \cite{AF_Extra,Altarelli_Lectures,MPT_SO10xA4,CDGG_Extra} and the flavour group which has been used is $A_4$. The mass matrices appear as
    \beq
    m_D\propto\unity\qquad\qquad\text{and}\qquad\qquad
    M_R\propto\left(
      \begin{array}{ccc}
        A & 0 & 0 \\
        0 & A & B \\
        0 & B & A \\
      \end{array}
    \right)\;.
    \eeq
    The charged leptons need to be rotated in diagonal form, and the main result is that the lepton mixing matrix is exactly the TB scheme.

  \item
    The second pattern \cite{BH_Geometric,HKV_A4,He_Renorm} is similar to the previous one and it still originates in an $A_4$ context. The mass matrices are the following:
    \beq
    m_D\propto\unity\qquad\qquad\text{and}\qquad\qquad
    M_R\propto\left(
      \begin{array}{ccc}
        A & 0 & B \\
        0 & A & 0 \\
        B & 0 & A \\
      \end{array}
    \right)\;.
    \eeq
    In the basis of diagonal charged leptons, we obtain the TB pattern for the lepton mixing matrix.

  \item
    The third pattern \cite{Ma_TBMandSUSYwithA4} is also similar to the first one. Once again it is based on the $A_4$ discrete symmetry. The mass matrices are given by
    \beq
    m_D\propto\unity\qquad\qquad\text{and}\qquad\qquad
    M_R\propto\left(
      \begin{array}{ccc}
        C & 0 & 0 \\
        0 & A & B \\
        0 & B & A \\
      \end{array}
    \right)\;.
    \eeq
    Like in the previous cases, when going to the basis of diagonal charged leptons it is easy to see that the lepton mixing matrix is the TB pattern.
\end{enumerate}

For all three patterns, it is trivial to see that $m_D^{R\,\dag} m_D^R$ is proportional to the identity matrix.


\subsubsection*{f)~~$\mathbf{(i,m)=(1,3)}$}

This case has been studied in two distinct patterns.

\begin{enumerate}
  \item
    In \cite{Yin_A4,HMV_Nie} a flavour model based on the $A_4$ group has been provided. The model is extremely similar to the previous case, of \cite{AF_Extra,Altarelli_Lectures,MPT_SO10xA4,CDGG_Extra}, where the structures of the Dirac and the Majorana mass matrices are exchanged:
    \beq
    m_D\propto\left(
      \begin{array}{ccc}
        A & 0 & 0 \\
        0 & A & B \\
        0 & B & A \\
      \end{array}
    \right)\qquad\qquad\text{and}\qquad\qquad
    M_R\propto\unity\,.
    \eeq
    In the basis of diagonal charged leptons the light neutrino mass matrix is diagonalised by the TB scheme and the product $m_D^\dag m_D$ is real.

  \item
    The second pattern has been presented in \cite{BFRS_A4Lepto} and it is similar to that of \cite{BH_Geometric,HKV_A4,He_Renorm}, discussed in the previous case, exchanging the structure of the Dirac and the Majorana mass matrices:
    \beq
    m_D\propto \left(
    \begin{array}{ccc}
        A & 0 & B \\
        0 & A & 0 \\
        B & 0 & A \\
      \end{array}
    \right)\qquad\qquad\text{and}\qquad\qquad
    M_R\propto\unity\;.
    \eeq
    This result has been developed in the context of the $A_4$ flavour symmetry \footnote{We underline the presence of the same difficulty previously discussed in $(1,2)$ about \cite{CK_Form}, which can be naturally solved by using $S_4$ instead.}. The authors themselves have concluded that $m_D$ does not give rise to leptogenesis.
\end{enumerate}

To conclude, each pattern in each case agrees with our model-independent result. Exact flavour symmetry imposed TB in type I see-saw leads to vanishing CP-asymmetry (the off-diagonal entries of $m_D^{R\,\dag} m_D^R$ are either trivially zero or real).

%
%

\section{Model dependent perturbations}
\label{sec:A4_2}

We concluded section \ref{exTBFS} with the observation that by
assuming general perturbations to the TB matrix obtained with an
underlying flavour symmetry there are no correlations between low and
high-energy scale CP violation parameters. This result was derived
from \eq{genper} where it can be seen that the number of free parameters governing the perturbations is quite large and thus no correlation can be expected. In the context of specific flavour models it is possible that the TB
scheme is perturbed by a small number of corrections, and in
this interesting case correlations between low-energy scale
observables and the CP-violating asymmetry may be established.

In this section we consider a supersymmetric model based on the $G_f=A_4\times Z_3\times Z_4$ discrete flavour symmetry \cite{Lin_Lepto}, which for our purposes is attractive due to its elegance and predictivity. The relevant NLO corrections appear only in the Dirac mass and can be parametrised in terms of only 3 complex parameters. Neutrino masses are induced only through type I see-saw so the results from section \ref{exTBFS} hold - in fact we have considered it explicitly in section \ref{sec:MBtutu}, as one of the models with the first pattern of class $(2,1)$.

The three factors in $G_f$ play different roles. The spontaneous breaking of $A_4$ is directly responsible for the TB mixing. The $Z_3\times Z_4$ factor avoids large mixing effects between the flavons that give masses to the charged leptons and those giving masses to neutrinos, and it is also responsible for the hierarchy among charged fermion masses.
The flavour symmetry breaking sector of the model includes the scalar superfields $\varphi_T$, $\xi'$, $\varphi_S$, $\xi$ and $\zeta$. The transformation properties of the lepton superfields $L$, $e^c$, $\mu^c$, $\tau^c$, of the electroweak scalar doublets $H^u$ and $H^d$ and of the flavon superfields are reproduced in table \ref{tab2} for ease of reference.

\begin{table}[h!]
\begin{center}
\begin{tabular}{|c|ccccccc|ccccc|}
\hline
&&&&&&&&&&&&\\[-3mm]
& $L$ & $e^c$ & $\mu^c$ & $\tau^c$ & $N^c$ & $H^u$ & $H^d$ & $\varphi_T$ & $\xi'$ & $\varphi_S$ & $\xi$ & $\zeta$ \\[3mm]
\hline
&&&&&&&&&&&&\\[-3mm]
$A_4$ & $3$ & $1$ & $1$ & $1$ & $3$ & $1$ & $1$ & $3$ & $1'$ & $3$ & $1$ & $1$ \\[3mm]
$Z_3$ & $1$ & $1$ & $1$ & $1$ & $\omega$ & $1$ & $1$ & $1$ & $1$ & $\omega$ &  $\omega$ &  $\omega^2$ \\[3mm]
$Z_4$ & $1$ & $-i$ & $-1$ & $1$ & $1$ & $1$ & $-i$ & $i$ & $i$ & $1$ & $1$ & $1$ \\[3mm]
\hline
\end{tabular}
\caption{Matter and scalar content of the model and their transformation properties under $G_f$ \cite{Lin_Lepto}.}
\label{tab2}
\end{center}
\end{table}

We present the Yukawa superpotential of the model as an expansions in $1/\Lambda$, where $\Lambda$ is the cut-off of the theory: the LO
terms are given by
\bea
&&\ba{rcl}
\mathcal{W}_\ell&=&\dfrac{1}{\Lambda} y_\tau \left(L\varphi_T\right)\tau^cH^d+ \\[3mm]
\\[-3mm]
&+&\dfrac{1}{\Lambda^2} y_\mu^{(1)} \left(L\varphi_T\right)''\xi'\mu^cH^d+\dfrac{1}{\Lambda^2} y_\mu^{(2)} \left(L\varphi_T\varphi_T\right)\mu^cH^d+ \\[3mm]
\\[-3mm]
&+&\dfrac{1}{\Lambda^3} y_e^{(1)} \left(L\varphi_T\right)'\left(\xi'\right)^2e^cH^d+\dfrac{1}{\Lambda^3} y_e^{(2)} \left(L\varphi_T\varphi_T\right)''\xi'e^cH^d+ \\[3mm]
\\[-3mm]
&+&\dfrac{1}{\Lambda^3} y_e^{(3)} \left(L\varphi_T\varphi_T\varphi_T\right)e^cH^d\;,
\ea\\[5mm]
&&\hspace{3mm}\mathcal{W}_\nu=-\dfrac{1}{\Lambda} y \left(LN^c\right)\zeta H^u+ x_a\left(N^cN^c\right)\xi+ x_b\left(N^cN^c\varphi_S\right)\;,
\eea
where $(\ldots)$, $(\ldots)'$ and $(\ldots)''$ stand for the contraction in the representations $1$, $1'$ and $1''$ of $A_4$, respectively.

The flavon superfields acquire the following VEVs:
\beq \ba{lllll}
\vev{\varphi_T}= \left(
                \begin{array}{c}
                  0 \\
                  v_T \\
                  0 \\
                \end{array}
              \right)\;, &
\vev{\xi'}= u'\;, &
\vev{\varphi_S}=\left(
                \begin{array}{c}
                    v_S \\
                    v_S \\
                    v_S \\
                  \end{array}
                \right)\;,&
\vev{\xi}= u\;, &
\vev{\zeta}= w\;,
\ea
\eeq
where $v_T$, $u'$, $v_S$, $u$ and $w$ are the small symmetry breaking parameters of the theory. This pattern of VEVs guarantees that the lepton mixing is approximately TB. It is possible to align these VEVs in a natural way, as the result of the minimisation of the scalar potential \cite{Lin_Lepto}: we underline that the symmetry content prevents any deviations from this pattern at NLO and allows the order of magnitude relations between parameters $v_T \sim u'$ and $v_S \sim u \sim w$, assuming at most a mild hierarchy among the two sets.

The charged lepton mass matrix can be approximately written as
\bea
m_\ell &=&
 \left(
  \begin{array}{ccc} \sim\frac{v_T^3}{\Lambda^3} & 0 & 0 \\
    0 & \sim\frac{v_T^2}{\Lambda^2} & 0 \\
    0 & 0 & \sim\frac{v_T}{\Lambda}
  \end{array}
\right)v^d\,.
\eea
where $v^d=\vev{H^d}$. A lower bound on the parameters $v_T/\Lambda$ can be fixed by the requirement that the $\tau$ Yukawa coupling $y_\tau$ does not become too large, and we can estimate it as
\beq
\dfrac{v_T}{\Lambda}=\dfrac{\tan\beta}{y_\tau} \dfrac{\sqrt{2} m_\tau}{v} \approx 0.01 \dfrac{\tan\beta}{y_\tau}
\label{tanb&u&yt}
\eeq
where $v\approx 246$ GeV and $\tan\beta=\vev{H^u}/\vev{H^d}$. Taking $m_\tau=(1776.84 \pm 0.17)$ MeV and requesting $|y_\tau|<3$, we find a lower limit on $v_T/\Lambda$ of $0.007$ for $\tan\beta=2$, the smallest value we consider.

The neutrino mass matrix gets contributions from the type I see-saw according to \eq{eq:lightNeu-MM}. We have:
\bea
\label{neutLO}
m_D=\left(
        \begin{array}{ccc}
        1 & 0 & 0 \\
        0 & 0 & 1  \\
        0 & 1 & 0
        \end{array}  \right)\dfrac{y\,w\,v^u}{\Lambda}\,,
  &\quad&
  M_R= \left(
        \begin{array}{ccc}
        b+2d & -d & -d \\
        -d & 2d & b-d \\
        -d & b-d & 2d
        \end{array} \right)u\,,
\eea
with $v^u=\vev{H^u}$, $b\equiv2x_a$ and $d\equiv2x_bv_S/u$. The mass matrices $M_R$ and $m_D$ are $\mu$-$\tau$ symmetric and
satisfy the conditions in \eq{cond1}. Accordingly, $M_R$ and $m_\nu$ are diagonalised by the TB mixing matrix, giving as eigenvalues $M_1=|b+3d|$, $M_2=|b|$, $M_3=|b-3d|$ and $m_i=(y\,w\,v^u)^2/(\Lambda^2M_i)$. We already mentioned in section \ref{sec:MBtutu} that the same mass matrices are present in \cite{FHLM_Tp,AFL_Orbifold,AF_Modular,AF_Extra}. The phenomenology has already been studied in \cite{BMM_SS}, so we summarise here the main results and refer to \cite{BMM_SS} for the details. The model can explain both Normal and Inverse Hierarchy (NH and IH) and features lower bounds on the mass of the lightest neutrino: in particular for the NH the lightest neutrino mass has a well defined and narrow range of values between $4.4$ meV and $7.3$ meV.

In order to estimate the parameter $\epsilon_{N_\alpha}$, we write the Dirac mass matrix in the basis of diagonal and real RH neutrinos:
\beq
m_D^R=m_DU_{TB}D'\,,
\eeq
where $D'=\diag(e^{i\phi_1/2},\,e^{i\phi_2/2},\,e^{i\phi_3/2})$ and $\phi_\al$ are the phases of $b+3d$, $b$, $b-3d$ respectively (the eigenvalues of $M_R$).
As was mentioned in section \ref{sec:MBtutu} the product $m_D^{R\dag} m_D^R$ is a diagonal matrix and therefore $\epsilon_{N_\alpha}=0$, in perfect agreement with our model-independent proof in section \ref{exTBFS}.

A non-vanishing CP-violating asymmetry can be obtained at NLO when the TB mixing is perturbed. In this
model the additional discrete Abelian symmetries $Z_3\times Z_4$ only admit NLO corrections to the Dirac terms. We do not consider terms whose contributions can be reabsorbed in a redefinition of the LO parameters, focusing only on terms that lead to deviations in the mixing angles:
\beq
-\mathcal{W}^{NLO}_\nu=\dfrac{1}{\Lambda} y_1 \left(LN^c\right)'\left(\varphi_S\varphi_S\right)''H^u+ \dfrac{1}{\Lambda} y_2 \left(LN^c\right)''\left(\varphi_S\varphi_S\right)'H^u+\dfrac{1}{\Lambda} y_3 \left(\left(LN^c\right)_A\varphi_S\right)\xi H^u\,,
\eeq
where $(\ldots)_A$ refers to the asymmetric contraction of the triplet representation. The deviations to $m_D$ can be written as
\beq
\label{md1}
m_D^{(1)}=
\left(
  \begin{array}{ccc}
    0 & y_1+y_3 & y_2-y_3 \\
    y_1-y_3 & y_2 & y_3 \\
    y_2+y_3 & -y_3 & y_1 \\
  \end{array}
\right)v^u\dfrac{v_S^2}{\Lambda^2}\,,
\eeq
where $y_3$ accounts for the ratio $u/v_S$.
Note that this correction is of Tri-maximal type \cite{Harrison:1999cf, Grimus:2008tt}. As the LO starts out as TB, and TB is also a (special) case of Tri-maximal, the perturbed model fits within that mixing scheme.
It is important to clarify that the general Tri-maximal scheme is explicitly out of the validity of the model-independent proof presented in section \ref{exTBFS} - while particular cases of Tri-maximal have mixing matrices independent of the masses (obviously this is the case for TB), in general it is possible to a Tri-maximal case where mixing angles depend on the masses. The perturbed model considered here is one such case, and as we will see it can admit viable leptogenesis.
Including \eq{md1}, the TB mixing receives small perturbations according to $U_\nu=U_{TB}\delta U$, where only the element $(\delta U)_{13}$ is relevant. Parametrising this term as:
\beq
(\delta U)_{13}=\sqrt{\dfrac{3}{2}}\sin\theta_{13}e^{i \delta }\sim\mathcal{O}\left(\dfrac{v_S}{\Lambda}\right)\,,
\eeq
where $\delta$ is the CP-violating Dirac phase in the standard parametrisation of the lepton mixing matrix, we write the other two mixing angles at NLO as
\beq
\label{expang}
\sin^2\theta_{23}=\dfrac{1}{2}(1+\sqrt2\cos \delta \sin\theta_{13}) \qquad\qquad\sin^2\theta_{12}=\dfrac{1}{3}(1+\sin^2\theta_{13})\,.
\eeq

\begin{figure}[ht!]
 \centering
\includegraphics[width=7.8cm]{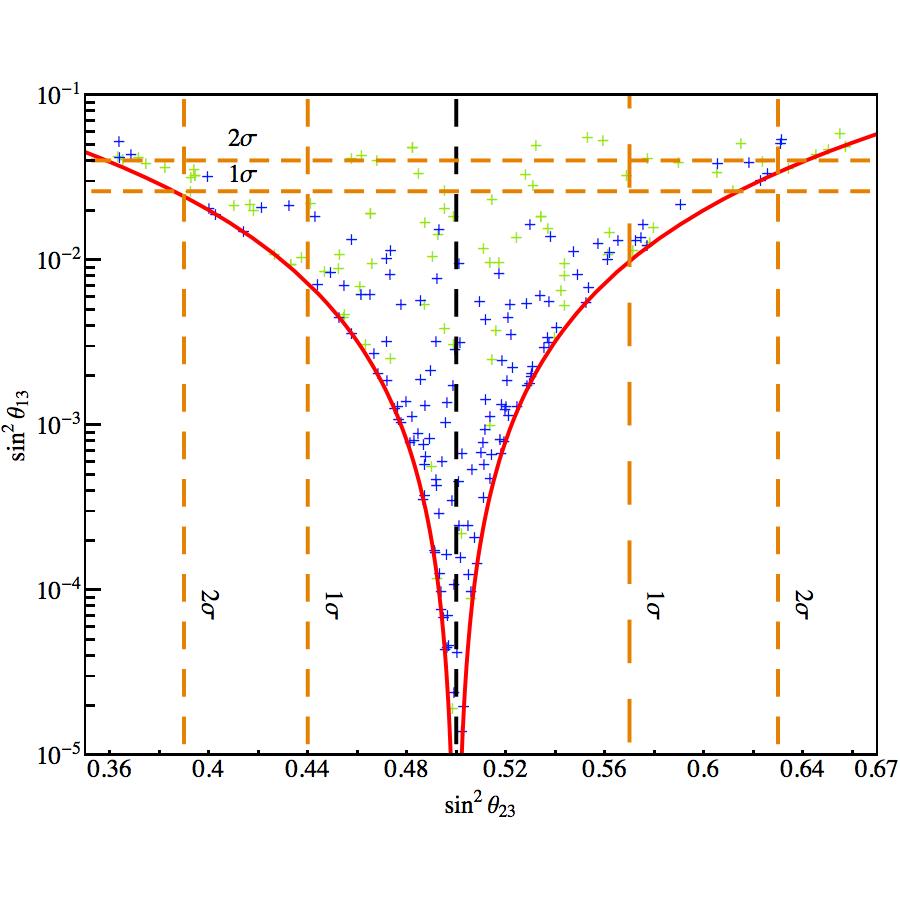}
 \includegraphics[width=7.8cm]{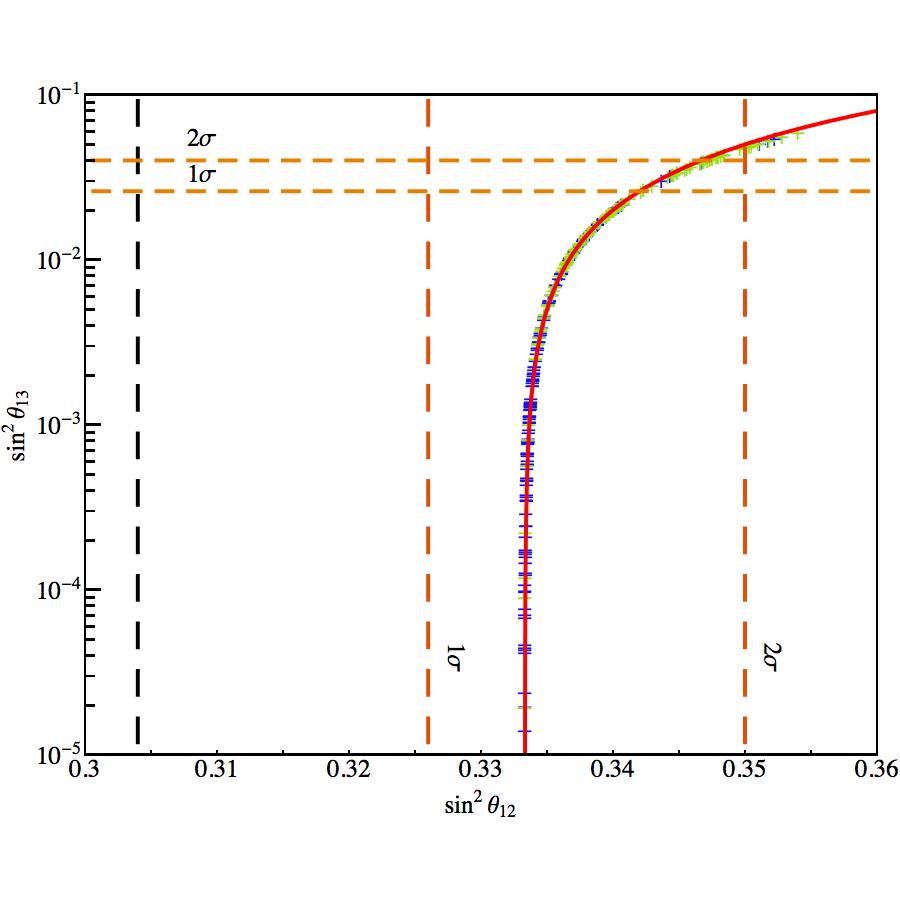}
 \caption{Correlation between $\sin^2\theta_{13}$ and  $\sin^2\theta_{23}$
   (left panel) and $\sin^2\theta_{12}$ (right panel). Each panel compares the analytical approximations given in eqs. (\ref{expang}) (red lines) with the numerical results. The green and blue points correspond to the IH and NH neutrino spectrum respectively. For the analytical expressions of $\sin\theta_{23}^2$ we have fixed the CP-Dirac phase $\delta$ at 0. In the numerical analysis the ratio $v_{S}/\Lambda\sim w/\Lambda$ has been taken in the window bounded  from  the constraints arising by  $y_\tau$ and $\sin^2\theta_{12} $, that is  $0.007<v_S/\Lambda<0.23$. The value of $\tan\beta$ spreads between $2$ and $50$, while all the other free parameters, $y$, $y_1$, $y_2$ and $y_3$, are treated as random numbers with absolute value between $0.1$ and $2$. The horizontal dashed orange lines corresponds to the bounds at $1$ and $2~\sigma$ level, respectively, for $\sin^2\theta_{13}$, the vertical dashed black line corresponds to the central values of $\sin^2\theta_{23}$ and $\sin^2\theta_{12}$ while the vertical dashed orange lines to their bounds at $1$ and $2~\sigma$ level, respectively. The plots are cut in correspondence of the $3~\sigma$ level bound for $\sin^2\theta_{13}, \sin^2 \theta_{23},\sin^2\theta_{12}$.}
 \label{fig:e13-23,12}
\end{figure}

In figure \ref{fig:e13-23,12}, we plot eqs. (\ref{expang}) in red (with $\delta=0$) and plot the results of a numerical analysis in green and blue points corresponding to the IH and NH neutrino spectrum respectively (in which we take $v_{S}/\Lambda\sim w/\Lambda=0.007\div0.23$, $\tan\beta=2\div50$ and we treat $y$, $y_1$, $y_2$ and $y_3$ as random numbers with modulus between $0.1$ and $2$).

We expect that NNLO corrections affect these relations: we estimate that NNLO perturbations will be of the order of $\sin^2\theta_{13}$ and therefore $\sin^2\theta_{12}$ will receive non-negligible corrections.

We can impose an upper bound on $v_S/\Lambda$ by requiring that the correction to the TB value of $\sin^2\theta_{12}$ does not take it outside the experimental $3\sigma$ range: the maximal allowed deviation from the TB value is $0.05$ and from there we impose the bound $v_S/\Lambda<\mathcal{O}(0.23)$.

We consider now $m_D^{R\prime}$ (the NLO Dirac neutrino mass matrix in the basis of diagonal and real RH neutrinos). We can write:
\beq
\label{expmDRp}
m_D^{R\prime}=m_D^R+m_D^{(1)}U_{TB}D'\,,
\eeq
and calculate the relevant product for leptogenesis, $m_D^{R\prime\dag}m_D^{R\prime}$, keeping only the first terms in the expansion in the small parameter $v_S^2/\Lambda^2$:
\beq
\label{md2}
m_D^{R\prime\dag}m_D^{R\prime}=m_D^{R\dag}m_D^R+ \left(D^*U_{TB}^Tm_D^{(1)\dag}m_D^R+h.c.\right)\,,
\eeq
where in the second term the only off-diagonal entries are the $13$ and $31$ ones. In the case of IH spectrum for the light neutrinos, the lightest RH neutrino is $N_2$: in this case the summation in the numerator of eq. (\ref{eq:cp-asymm}) does not contain the term $13$ and therefore $\epsilon_{N_2}$ is vanishing also at NLO. This, however, does not mean leptogenesis can not be realized in this case. Since there is only a mild hierarchy between $N_2$, $N_1$ and $N_3$ and neither $\epsilon_{N_1}$ nor $\epsilon_{N_3}$ vanishes, leptogenesis will proceed through $N_{1,3}$ dynamics. In the NH case the RH neutrino mass spectrum follows the hierarchy $M_{N_3}<M_{N_2}<M_{N_1}$. There is a mild hierarchy between $N_3$ and $N_2$ while the hierarchy between $N_3$ and $N_1$ is large (around a factor 9). Consequently, the lepton asymmetry generated in $N_1$ decays will be, in general, erased by the lepton number violating interactions of $N_3$. Only $N_3$ dynamics becomes relevant for the generation of a lepton asymmetry in this case. Note that if the hierarchy between $N_1$ and $N_3$ decreases (as could be in the case of a quasi-degenerate spectrum), so it becomes mild, $N_1$ dynamics should be taken into account. Henceforth, for simplicity, we will consider only the NH case for which, according to eq. (\ref{eq:cp-asymm}), the CP-violating parameter $\epsilon_{N_3}$ can be written as
\beq
\label{expep}
\epsilon_{N_3}=\dfrac{1}{8\pi}\dfrac{1}{v_u^2\left(m_D^{R\,\dagger} \;m_D^R\right)_{11}}\mathbb{I}\mbox{m}\left[\left(\left(m_D^{R\,\dagger} \;m_D^R\right)_{13}\right)^2\right]f\left(\dfrac{M_1^2}{M_3^3}\right)\,.
\eeq

In the following figures we show a series of scatter plots related to the predictions of the model and the connections among low-energy observables and $\epsilon_{N_3}$. The (blue) points correspond only to the NH neutrino spectrum (in which we take $v_{S}/\Lambda\sim w/\Lambda=0.007\div0.23$, $\tan\beta=2\div50$ and we treat $y$, $y_1$, $y_2$ and $y_3$ as random numbers with modulus between $0.1$ and $2$). Red lines correspond to analytical results.

\begin{figure}[ht!]
  \centering
\includegraphics[width=7.8cm]{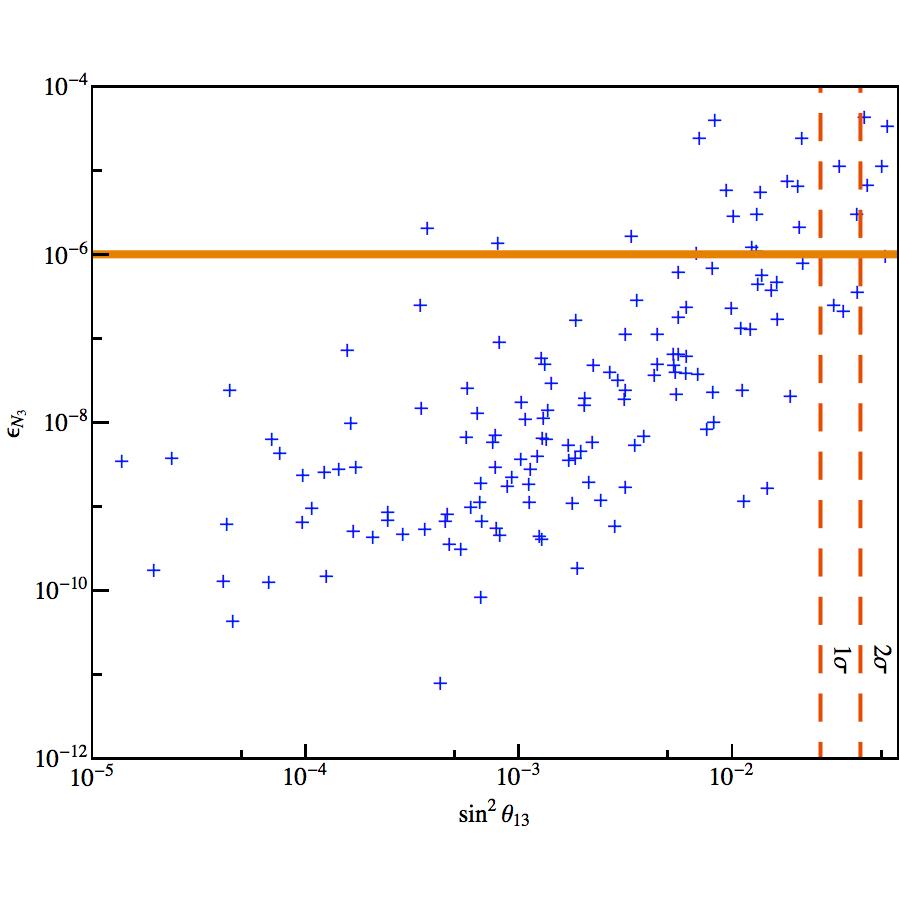}
\includegraphics[width=7.8cm]{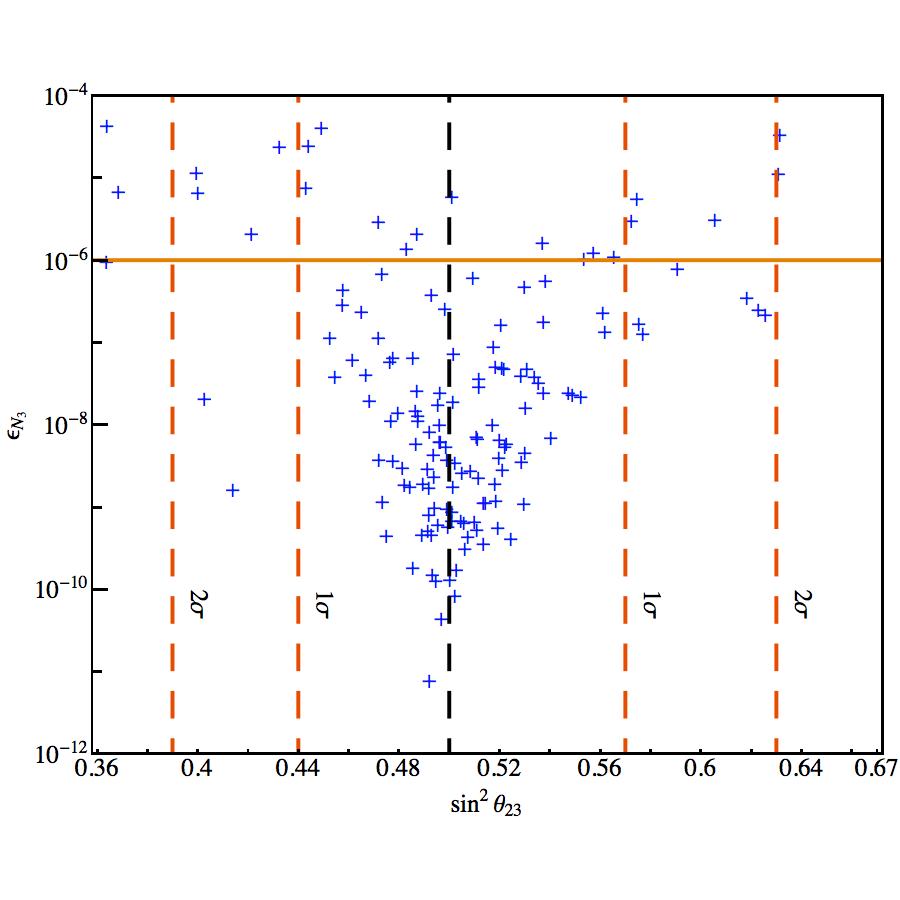}
\includegraphics[width=7.8cm]{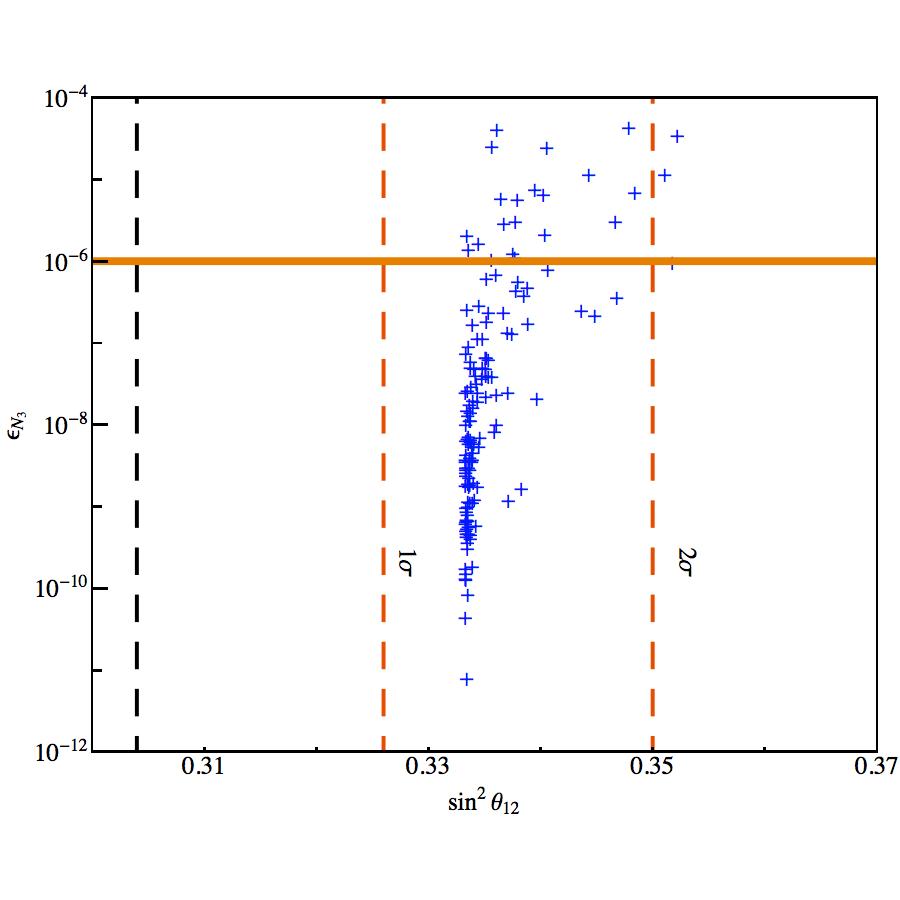} \caption{Correlation between $\epsilon_{N_3}$ and $\sin^2\theta_{13}$ (first row,  left panel), $\sin^2\theta_{23}$ (first row, right panel), $\sin^2\theta_{12}$ (second  row). The horizontal orange line corresponds to $\epsilon_{N_3}\sim 10^{-6}$, the vertical orange lines correspond to the bounds at $1$ and $2~\sigma$ level  of $\sin^2\theta_{13}$, $\sin^2\theta_{23}$ and $\sin^2\theta_{12}$ respectively while the vertical dashed black lines correspond to the central values of $\sin^2\theta_{23}$ and $\sin^2\theta_{12}$. The range presented in the plots covers the $3~\sigma$ level bounds.}
 \label{fig:eN-13-23}
\end{figure}

In figure~\ref{fig:eN-13-23} we show all the correlations between the CP asymmetry parameter $\epsilon_{N_3}$ and the lepton mixing angles (expressed as $\sin^2\theta_{13}$, $\sin^2 \theta_{23}$ and $\sin^2 \theta_{12}$). As expected by comparing \eq{expep} with \eq{expang}, $\epsilon_{N_3}$ is correlated to all low-energy mixing angles.

\begin{figure}[ht!]
 \centering
\includegraphics[width=8cm]{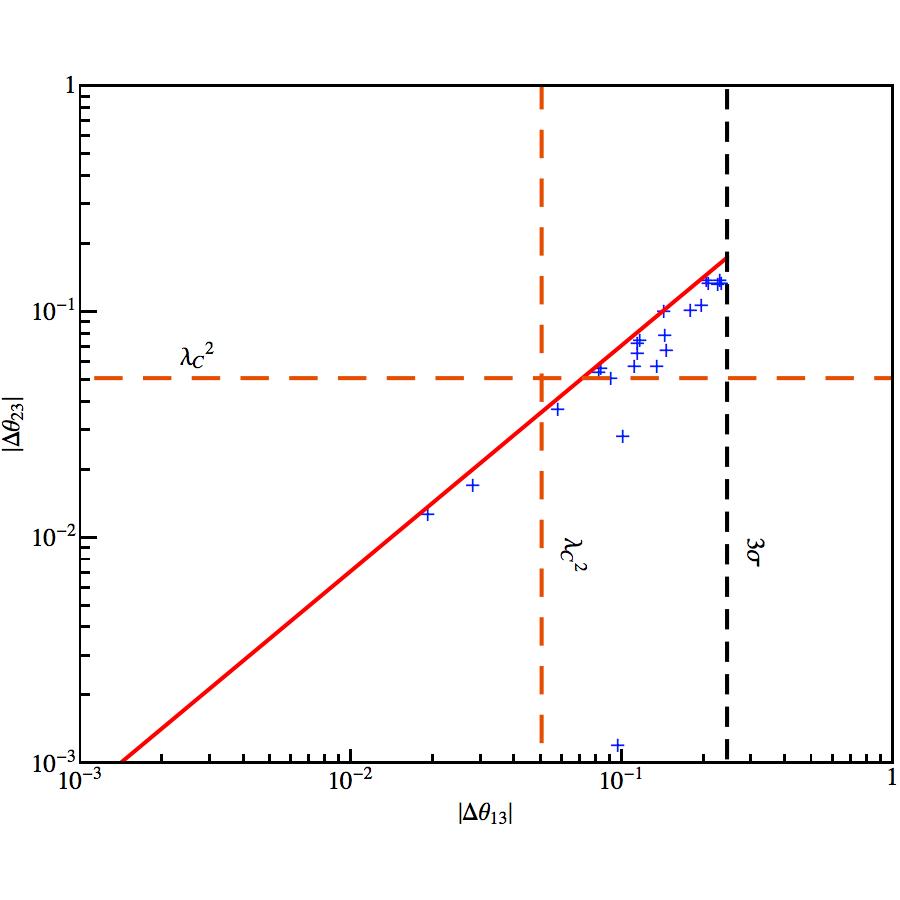}
\includegraphics[width=8cm]{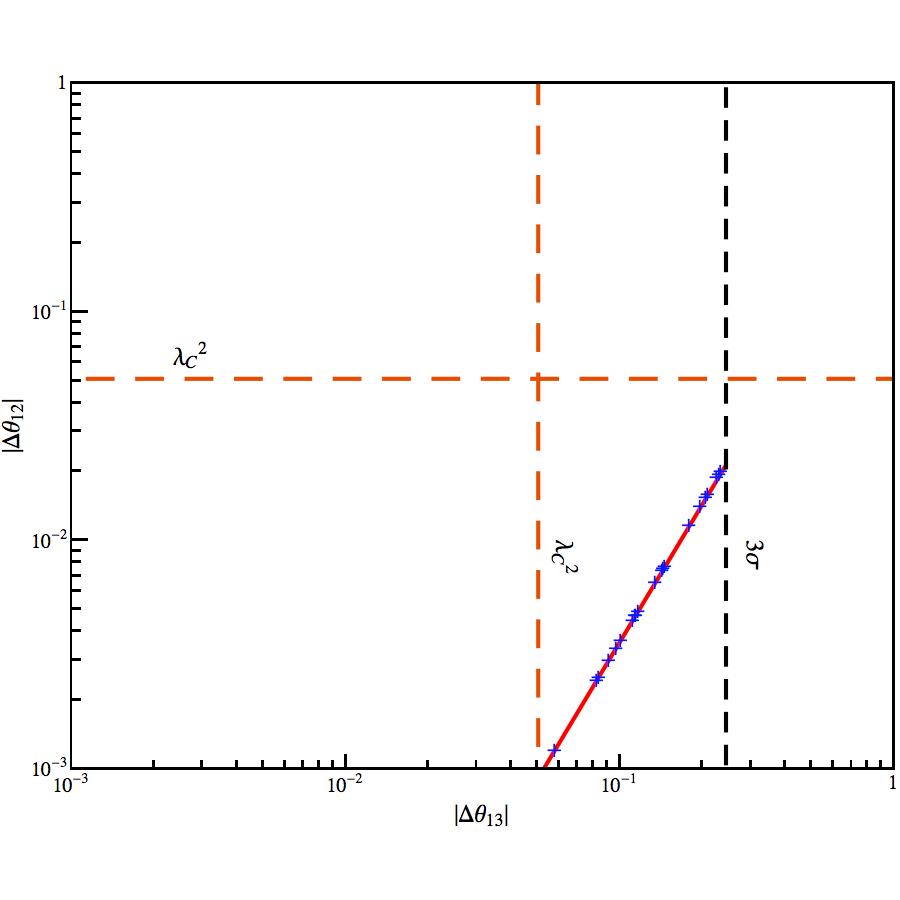}
\caption{Correlation between the deviation of $\theta_{23}$  (left panel) and $\theta_{12}$ (right panel)  with
 $\theta_{13}$ from their TB mixing values for the points that satisfy
  $\epsilon_{N_3} \geq 10^{-6}$. The red lines represent the analytical results from eqs. (\ref{expang}). The vertical and horizontal orange lines corresponds to $|\Delta\theta_{13}|,|\Delta\theta_{23}|, |\Delta\theta_{12}|\sim\lambda_C^2$
  with $\lambda_C$ the Cabibbo angle.  }
  \label{fig:dev}
\end{figure}

The same information is contained in figure~\ref{fig:dev} where we show the deviations of $\theta_{13},\theta_{12},\theta_{23}$ from the respective TB value for the points that reach  the necessary amount of CP asymmetry. The deviations are compared to the reference value $\lambda_C^2\sim 0.05$, where $\lambda_C$ is the Cabibbo angle. This comparison is particularly interesting because $\lambda^2_C$  is the typical  order of magnitude of the corrections to the TB mixing allowed by neutrino data fit  in models  based on flavour symmetries and predicting TB mixing (in particular, this is a natural consequence in classes of flavour models that include GUTs \cite{continuous}). Our numerical analysis shows indeed that the order of magnitude of  $|\Delta\theta_{23}|,|\Delta\theta_{13}|$ is close to $\lambda_C^2$ while  $|\Delta\theta_{12}|$ tends to be smaller. We note that only a few points reach the necessary amount of CP asymmetry $\epsilon_{N_3}$ when the  deviations of the mixing angles from their TB values are smaller than $\lambda_C^2$.  By comparing the left and right panel of figure~\ref{fig:dev} we can get  the lower bound for $\sin^2\theta_{13}$ necessary to have  successful leptogenesis in this model, that is  $\sin^2\theta_{13}\sim 10^{-2}$.
Future experiments will further constrain (and possibly rule out leptogenesis within this model). Double Chooz will probe $\sin^2 ( 2\theta_{13})$ to $10^{-2}$ in the next five years and Triple Chooz will reach below that value \cite{Huber:2006vr}.

\begin{figure}[ht!]
 \centering
\includegraphics[width=7.8cm]{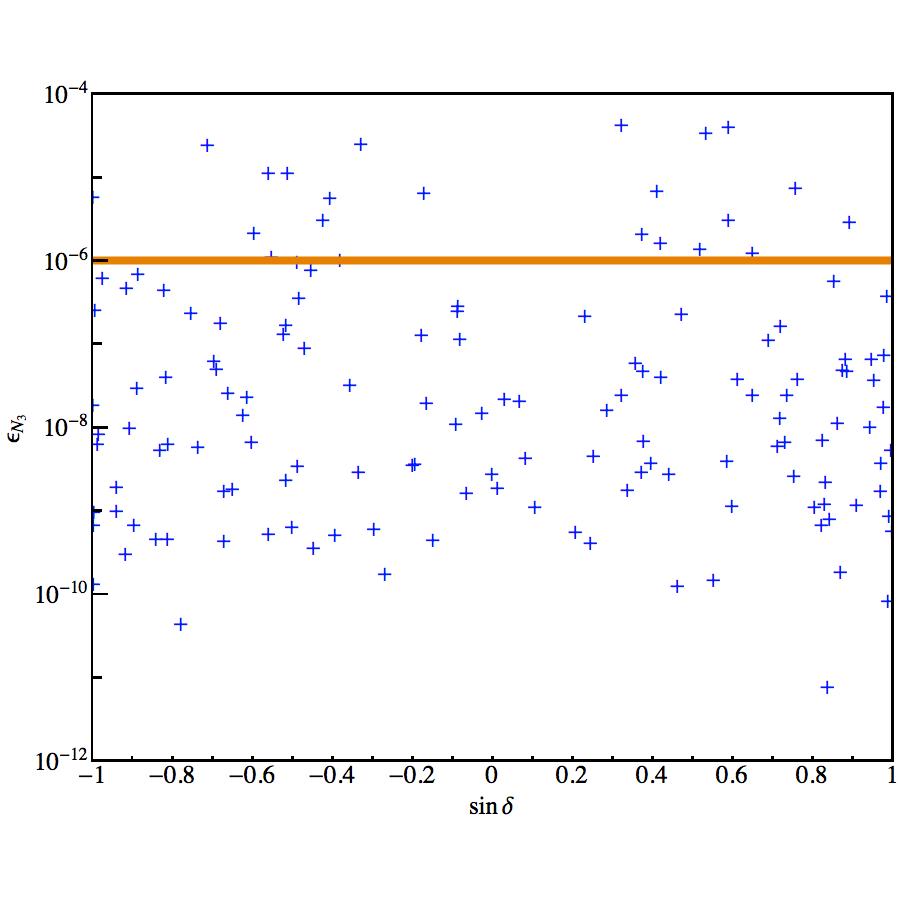}
\includegraphics[width=7.8cm]{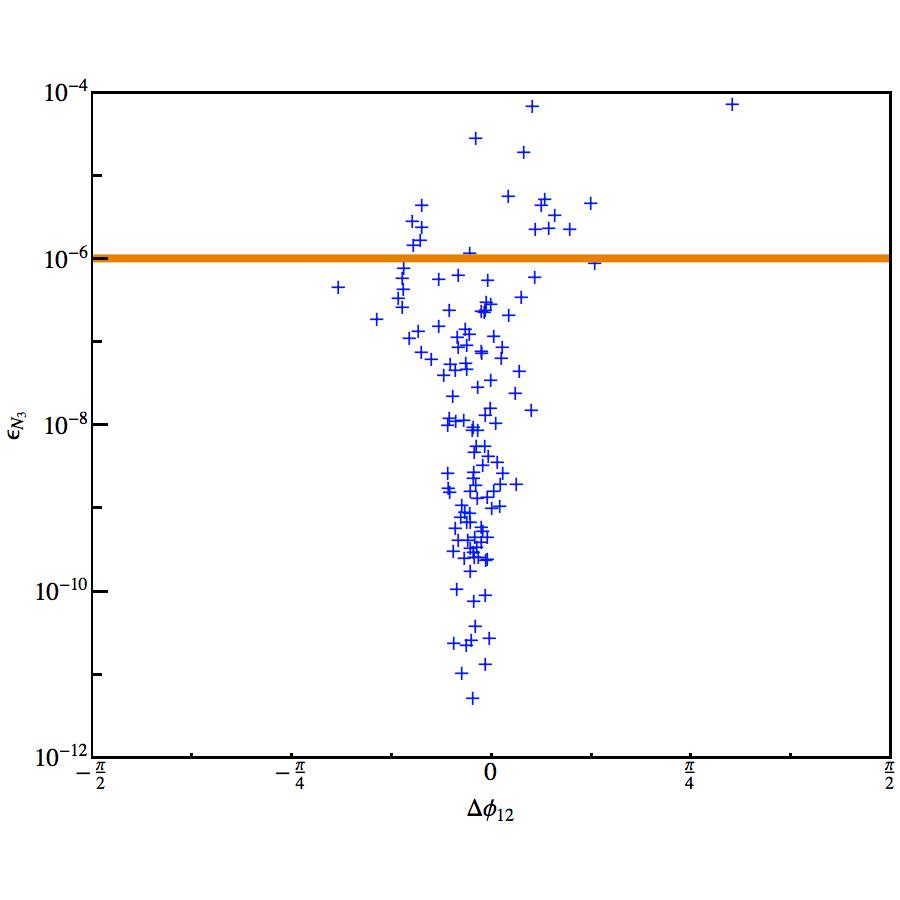}
\caption{The high energy CP violation asymmetry parameter
  $\epsilon_{N_3}$ versus the low energy lepton CP-Dirac phase $\delta$ and the difference of the light neutrino Majorana phases $\phi_1$ and $\phi_2$. In both plots the horizontal orange line corresponds to $\epsilon_{N_3}\sim 10^{-6}$.}
 \label{fig:del-lep}
\end{figure}

Finally figure~\ref{fig:del-lep} wants to investigate the possible correlations between the  high-energy CP asymmetry parameter $\epsilon_{N_3}$ and the low-energy CP-Dirac and Majorana  phases. The left panel of  figure~\ref{fig:del-lep} shows  that  the low energy  CP-Dirac phase  $\delta$ ($\delta$ in the plot)  is not correlated to $\epsilon_{N_3}$.  This result is not surprisingly: the phases that enter in $\epsilon_{N_3}$ are related to the phases  present in $M_R$, while $\delta$ arises by the  phases that appear in what we defined as $m_D^{(1)}$ in \eq{expmDRp}.  On the contrary the right panel of  figure~\ref{fig:del-lep} indicates that the difference between the Majorana phases $\phi_1$ and $\phi_2$ ($\Delta \phi_{12} $ in the plot) presents a correlation with
$\epsilon_{N_3}$. The reason is very simple: at LO the phenomenological analysis of this model shows that the NH spectrum can be reproduced only if  $\Delta \phi^0_{12} $ is small.  Moreover at LO, $\Delta\phi^0_{12}$  coincides with the corresponding Majorana phase difference  $\Delta \phi_{12}^R$  of the right handed neutrinos. By perturbing the neutrino Dirac mass matrix we introduce new arbitrary phases that can vary in all the interval $(0,2 \pi)$. However the NLO contributions are responsible both of deviating the lepton mixing angles  by the TB values and  slightly modifying the neutrino spectrum, in the range allowed by the data fit. This means that in general at NLO the neutrino (complex) mass eigenvalues are given by
\beq
m_i \sim m_i^0 +\delta m_i\,,
\eeq
where $\delta m_i$ are complex parameters and $m_i^0$  the neutrino mass eigenvalues at LO. Requiring now that $\Delta m^2_{12}$ is still in the range indicated for $\Delta m^2_{sol}$ we have that $\delta m\sim |\delta m_{1,2}| \sim 10^{-3}$ eV for  $|m_1^0|,|m_2^0|\sim \mathcal{O}(\sqrt{\Delta m^2_{sol}})$. A straight computation shows then that the Majorana phase $\Delta \phi_{12}$ satisfies
\beq
\tan  \Delta \phi_{12}\sim  \tan  \Delta \phi^0_{12}+ \alpha \frac{\delta m}{ \mathcal{O}(\sqrt{\Delta m^2_{sol}})}\,,
\eeq
where $\Delta \phi^0_{12}$ is the LO  Majorana phase difference and $\alpha\in (0,1)$ a parameter that takes into account that the $\delta m_{1,2}$  phases run into the interval $(0, 2 \pi)$. We can estimate the maximal deviation of $\Delta\phi_{12}$ by its LO value getting
\beq
\Delta\phi_{12}- \Delta\phi^0_{12} \sim \frac{\pi}{10}\,.
\eeq
Notice that the left panel of fig.~\ref{fig:del-lep} shows that the majority of all the points are indeed inside the interval $(-\pi/8,\pi/8)$ in perfect agreement with our  analytical results  for a small LO  $\Delta\phi^0_{12}\leq0.1$.

In summary our analysis shows that in the model considered it is possible to obtain correlations between low-energy observables  and the high-energy CP-violating parameter, but it confirms that in general no correlation is present between  high and low-energy CP-violating parameters (or in the case of the Majorana phases, negligible correlation).


\section{Conclusion}
\label{sec:end}

In this paper we considered under rather general conditions the
possibility of links between low-energy observables and high-energy
parameters that are relevant for leptogenesis - in the most general case no such
connections can be recovered.

When assuming exact TB mixing independently of any specific
justification, we conclude that it is in general possible to obtain
leptogenesis. Constraining the situation to the case of type I see-saw
is insufficient to provide a link between the different type of
parameters.

In the main part of this work, we considered the more natural case where exact mixing patterns originate from any
flavour symmetry. We confirmed that the results of \cite{JM_A4Lepto} concerning TB mixing apply
to the case of unflavoured leptogenesis when there is only type I see-saw. We generalised this conclusion into a model-independent proof that is also valid for other flavour symmetry imposed mixing schemes if the mixing matrix consists purely of numbers - this includes Bi-maximal mixing, golden-ratio mixing, and some but not all cases of Tri-maximal mixing.
We emphasise that the proof does not hold when there are also other types of see-saw (such as type II): in models in which there is interplay between different see-saws, it is possible to have leptogenesis without lifting the exact (TB) pattern. These interesting cases shall be considered in detail in future work \cite{future}.
Still in the model-independent framework with only type I see-saw, we considered the most general NLO corrections that can lift TB mixing and how these corrections can enable leptogenesis.

From the model-independent proof we proceeded by considering several flavour symmetry models with exact TB mixing and only type I see-saw.
As expected, in all cases the specific conditions led to vanishing CP asymmetry.

Finally we studied a specific example in which a flavour symmetry model has deviations
from exact TB, which can then enable leptogenesis.  In
a general case there would be many parameters governing the deviation
from TB mixing and thus an interesting link between observable mixing
angles and leptogenesis may not even exist. We selected an example
where the deviation is parametrised such that it is possible to obtain relatively simple
analytical expressions relating the observable deviations from TB
angles to the CP asymmetry. There is a clear and strong correlation between deviations from
TB angles (such as the non-vanishing value of $\theta_{13}$) and the
value of the CP asymmetry parameter: particularly we note that if we insist in having viable leptogenesis in this model we can considerably constrain the allowed parameter space of the NH spectrum. Future experiments will probe the remaining parameter space.

To summarise, in order to have TB mixing scheme originating from a flavour symmetry and still have viable leptogenesis the model requires NLO corrections lifting the exact mixing, or alternatively it requires independent contributions to the CP asymmetries such as those that naturally arise from an interplay between different see-saws.


\section*{Acknowledgments}
We thank Yin Lin and Enrico Nardi for useful discussions.  The work of FB has been
partially supported by European Commission Contracts
MRTN-CT-2004-503369 and ILIAS/N6 RII3-CT-2004-506222 and by the
foundation for Fundamental Research of Matter (FOM) and the National
Organization for Scientific Research (NWO).
The work of IdMV was supported by FCT under the grant
SFRH/BPD/35919/2007.  The work of IdMV was partially supported by FCT
through the projects POCI/81919/2007, CERN/FP/83503/2008 and CFTP-FCT
UNIT 777 which are partially funded through POCTI (FEDER) and by the
Marie Curie RTN MRTN-CT-2006-035505.  LM recognises that this work has
been partly supported by the European Commission under contract
MRTN-CT-2004-503369 and MRTN-CT-2006-035505. The work of SM supported
by MEC-Valencia MEC grant FPA2008-00319/FPA, by European Commission
Contracts MRTN-CT-2004-503369 and ILIAS/N6 RII3-CT-2004-506222.

\section*{Note added in proof}
While completing this paper we received refs.
\cite{Bertuzzo:2009im,Hagedorn:2009jy}, where the interplay of flavour
symmetries and leptogenesis in the context of type I see-saw is also
discussed. Both papers consider specific models based on the discrete
group $A_4$. Furthermore, ref. \cite{Bertuzzo:2009im} discusses a
general model-independent approach which complements the distinct
model-independent proof we provide here.




\end{document}